# Visible-Infrared spectroscopy of ungrouped and rare meteorites brings further constraints on meteorite-asteroid connections.

L. Krämer Ruggiu[1], P. Beck[2], J. Gattacceca[1], J. Eschrig[2].

[1]Aix Marseille Univ, CNRS, IRD, INRAE, CEREGE, Aix-en-Provence, France

(kramer@cerege.fr)

[2]Univ. Grenoble Alpes, CNRS, IPAG, Grenoble, France

Corresponding author: kramer@cerege.fr

**Abstract**

The composition of asteroids gives crucial insights into the formation and evolution of the Solar System. Although spectral surveys and spacecraft missions provide information on small bodies, many important analyses can only be performed in terrestrial laboratories. Meteorites represent our main source of samples of extraterrestrial material. Determining the source asteroids of these meteorites is crucial to interpret their analyses in the broader context of the inner Solar System. For now, the total number of parent bodies represented in our meteorites collection is estimated to about 150 parent bodies, of which 50 parent bodies represented by the poorly studied ungrouped chondrites. Linking ungrouped meteorites to their parent bodies is thus crucial to significantly increase our knowledge of asteroids. To this end, the petrography of 25 ungrouped chondrites and rare meteorite groups was studied, allowing grouping into 6 petrographic groups based on texture, mineralogy, and aqueous and thermal parent body processing. Then, we acquired visible-near-infrared (VIS-NIR) reflectance spectroscopy data of those 25 meteorites, in order to compare them to ground-based telescopic observations of asteroids. The reflectance spectra of meteorites were obtained on powdered samples, as usually done for such studies, but also on raw samples and polished sections. With asteroids surfaces being more complex than fine-grained regolith (e.g., asteroid (101955) Bennu), in particular near-Earth asteroids, the use of raw samples is a necessary addition for investigating parent bodies. Our results showed that sample preparation influences the shape of the spectra, and thus asteroid spectral matching, especially for carbonaceous chondrites. Overall, the petrographic groups defined initially coincide with reflectance spectral groups, with only few exceptions. The meteorite spectra were then compared with reference end-member spectra of asteroids taxonomy. We matched the 25 studied meteorites to asteroids types, using a qualitative match of the shape of the spectra, as well as a quantitative comparison of spectral parameters (bands positions, bands depths and slopes at 1 and 2 µm). We define links between some of the studied ungrouped chondrites and asteroid types that had no meteorite connection proposed before, such as some very primitive type 3.00 ungrouped chondrites to B-type or Cg-type asteroids. We also matched metamorphosed ungrouped carbonaceous chondrites to S-complex asteroids, suggesting that this complex is not only composed of ordinary chondrites or primitive achondrites, as previously established, but may also host carbonaceous chondrites. Conversely, some ungrouped chondrites could not be matched to any known asteroid type, showing that those are potential samples from yet unidentified asteroid types.

# 1 Introduction

The knowledge of the spatial distribution and composition of Solar System small bodies (asteroids, Trojans, trans-Neptunian objects (TNOs) and comets) gives us insight into the formation and dynamical evolution of the Solar System. The composition of small bodies is mainly known from the analysis of the sunlight reflected from their surfaces. Indeed, the major minerals composing asteroids (silicates) have specific absorption bands in the visible (VIS) and Near Infrared (NIR) wavelengths. Spectral surveys have measured the reflectance of asteroids in the visible range of wavelengths (~0.4-1.1 µm) (e.g. Tholen, 1984; Zellner et al., 1985; Bus, 1999; Lazzaro et al., 2004) and spectra up to 2.5 µm are available for several hundreds of objects.

Using those spectra, asteroids have been grouped into different classes. The Tholen taxonomy (1984), based on the Eight Color Asteroids Survey data (ECAS, Zellner et al., 1985), defined seven major asteroid classes. Using the spectral diversity of the new CCD technology, Bus and Binzel (2002) proposed an improved taxonomy with three major complexes (S, C and X), composed of 12 classes, including 26 different types of asteroids. Both the Tholen and Bus taxonomies are based on Principal Component Analysis (PCA) of NIR spectral parameters. This taxonomy was then revised by DeMeo et al. (2009) based on spectra up to 2.5 μm, eliminating three classes, creating a new one, resulting in a total of 24 asteroid classes.

Spacecraft mission is another way to analyze asteroids, and collect information unobtainable from Earth, such as spatially resolved reflectance spectra, high resolution image, bulk densities, and magnetic field measurements. Those missions have so far visited only a dozen asteroids: (1) Ceres and (4) Vesta (Dawn, Krohn et al., 2018), (253) Mathilde and (433) Eros (NEAR Shoemaker, Prockter et al., 2002), (243) Ida and (951) Gaspra (Galileo, Johnson et al., 1992), (21) Lutetia and (2867) Steins (Rosetta, Glassmeier et al., 2007), (4179) Toutatis (Chang'e 2, Huang et al., 2013), (5535) Annefrank (Stardust, Duxbury et al., 2004), (9969) Braille (Deep Space 1, Buratti et al., 2004), (25143) Itokawa (Hayabusa, Tsuda et al., 2013) and (486958) Arrokoth (New Horizons, Young et al., 2008). During these missions, detailed elemental composition has only been obtained for a few of these bodies ((1) Ceres, (4) Vesta, (433) Eros, (25143) Itokawa) (Burbine, 2016). A third approach to study small bodies is through sample return missions. So far, one comet Wild 2 (Stardust mission) and two asteroids (25143) Itokawa (Hayabusa mission) and more recently (132173) Ryugu (Hayabusa2 mission) have been successfully sampled. One additional sample return mission, OSIRIX-Rex to asteroid (101955) Bennu is underway. The study of meteorites is another method to study samples of small bodies. Meteorite collections contain over 60000 meteorites collected and catalogued since the early 1800's (see the Meteoritical Bulletin Database, https://www.lpi.usra.edu/meteor/). Over 99.9% of those meteorites derive from asteroids, and only a small fraction from planetary bodies such as Mars and the Moon (Burbine et al., 2002). For chondrites, meteorites in the same group are thought to originate from the same primary parent body (Greenwood et al., 2020). Achondrites' classification is different and not only based on their parent bodies, but on petrography and chemistry, such that, for instance, meteorites originating from asteroid Vesta are separated into three groups (Howardites, Eucrites, Diogenites). The current classification comprises 50 groups of meteorites (Weisberg et al., 2006). In this study, we focus on chondrites, that can be classified in 3 major classes: Ordinary (O), Carbonaceous (C) and Enstatites (E); and 2 additional classes: Kakangari (K) and Rumuruti (R). Ordinary chondrites are the most abundant meteorite type, subdivided in three groups, H, L and LL. The carbonaceous chondrites are divided in several classes by petrographic and geochemical variations: CI, CM, CK, CV, CO, CR, CB, and CH (Weisberg et al., 2006; Krot et al., 2014). However, these chondrites groups represent only about half of the chondrite diversity. Indeed, some meteorites do not fit any of the groups and are called "ungrouped". Those ungrouped meteorites cannot be classified in one of the main groups due to non-matching isotopic compositions, different petrographic characteristics, and/or mineral compositions. According to the Meteoritical Bulletin database (accessed September 2020), most of the ungrouped chondrites are carbonaceous chondrites (55 of 86 ungrouped chondrites). These meteorites have been classified as C2-ung and C3-ung meteorites, or just C2 or C3 until recently. C2 and C3 designation should be exclusively reserved to meteorites whose classification is not detailed enough to exclude affinities with a known chondrite group, but C2 chondrites, for example, comprises real ungrouped chondrites (C2-ung) but also many CM2 chondrites.

As a result, it is not straightforward to evaluate the number of ungrouped chondrites, and hence the number of parent bodies that they sample. Based on oxygen isotopic composition considerations, Greenwood et al. (2020) proposed that ungrouped chondrites originate from a maximum of 50 different parent bodies; ungrouped achondrites derived from 16 distinct parent bodies; and ungrouped iron meteorites from a maximum of 50 parent bodies. Counting chondrites, achondrites and ungrouped meteorites, the total number of parent bodies represented in our collection is thus estimated to about 150 parent bodies (Gattacceca et al., 2020a). Based on oxygen isotopes studies, Greenwood et al. 2020 estimated the number of different parent bodies to 95-148 asteroids, with 15-20 parent bodies for chondrites, and 11-17 parent bodies for ungrouped chondrites. Grouped meteorites usually come from collisional families composed of multiple smaller objects, thus, our sampling of asteroids may be biased toward family members (e.g. DeMeo et al., 2013; Vernazza and Beck, 2016). Although meteorites are "float" rocks, i.e. collected outside of their geological context, studying ungrouped meteorites may enable to sample rare asteroid types that possibly did not produce collisional families.

Linking meteorites to asteroids is necessary to study the precise mineralogy of asteroids. The study of chemical and isotopic compositions of asteroids, being building block of planet formation, allow better understanding of composition and thermal gradients in the solar nebula. Manned space missions may want to search asteroids for important resources relatively rare on Earth (Kargel, 1994). Also, near-Earth asteroids (NEAs) can strike the Earth and any deflection or destruction strategy would need to include the mineralogy of those objects (Burbine et al., 2002).

A first approach to link meteorites to their origin in the Solar System is dynamical models, that find that most meteorites originate from the inner main belt (e.g., Granvik and Brown, 2018). More precisely, to link meteorites to their parent bodies is possible using automated camera networks monitoring the sky for meteoroids. The goal of these networks is to find the heliocentric orbits of meteoroids and ultimately their source regions in the Solar System (Granvik and Brown, 2018). The use of these networks enabled to link only about 25 meteorites to their parent bodies (Granvik and Brown, 2018).

The most used method to connect meteorite groups to parent bodies is finding spectral similarities in the visible and near-infrared (NIR) (e.g. Thomas and Binzel, 2010; Burbine, 2016 and references therein; Takir et al., 2019). Reflectance spectra of meteorites are measured in laboratory and compared to asteroid reflectance spectra acquired through telescopes (Takir et al., 2019, Potin et al., 2020). In addition, meteorites groups have distinctive spectra and absorptions in the spectral range used for asteroids reflectance spectroscopy (0.3-2.5 µm) (Gaffey, 1976). The comparison between meteorite and asteroid spectra is complicated as visible-near-infrared (VIS-NIR) spectroscopy can be influenced by other factors than mineralogy, such as space weathering (Chapman, 1996; Sasaki et al., 2001; Hapke, 2001), grain size (Johnson and Fanale, 1973), temperature (Singer and Roush, 1985; Hinrichs et al., 1999) or adsorbed water (Beck et al., 2010; Garenne et al., 2016; Takir et al., 2019). Despite these complications, links have been proposed between some asteroid complexes and the main meteorites groups (see review Table 2 in Greenwood et al., 2020), and most meteorites are linked to asteroids collisional families, such as HEDs (howardites, eucrites, diogenites) to the Vesta family (DeMeo et al., 2013).

Although the spectral properties of the main meteorite groups have been studied in details (e.g. Burbine et al., 2002; Beck et al., 2013; Burbine 2016; Vernazza et al., 2017; Takir et al., 2019; Eschrig et al., 2020), the spectral properties of the ungrouped chondrites have received much less attention, even though they represent a large fraction of the diversity of the meteoritic material available for study. In addition, only a few studies have focused on connecting ungrouped chondrites to asteroids. Indeed, Tagish Lake meteorite (C2-ung) was linked to either D- or T-type asteroids by Hiroi et al. (2001) and Hiroi and Hasegawa (2003), and Izawa et al. (2015) linked the meteorite to two NEAs ((326732) 2003 HB6 and (17274) 2000 LC16). Thus, the rest of the ungrouped meteorites, which represent various meteorites mineralogies, and thus potential parent bodies, are probable samples of other unlinked asteroids types.

In this paper, we study 25 meteorites (mostly chondrites), ungrouped or from rare groups, whose spectral properties have never been studied or at least not in detail. First, we analyzed the petrography of these meteorites to put some order in the hazy "ungrouped" designation. Then we acquired Infrared (IR) reflectance data on these 25 meteorites and compared these results with reference asteroids data. We discuss the limitations and parameters influencing the comparison of asteroid to meteorite spectra, and propose possible connections between some of the studied meteorites and potential asteroids parent bodies.

# 2 Samples and Methods

## 2.1 Samples

A suite of 25 meteorites was used in this study, with a focus on ungrouped chondrites. We studied 12 ungrouped chondrites, 2 anomalous chondrites and 11 meteorites from rare groups (Table 1). The ungrouped meteorites comprise one ungrouped ordinary chondrite (JaH 846), one ungrouped chondrite (Sierra Gorda 009), two ungrouped carbonaceous chondrites (Dhofar 2066 and North West Africa (NWA) 8781), and three C2-ung chondrites (NWA 5958, Acfer 094 and El Médano (EM) 100). We also studied three ungrouped carbonaceous chondrites of petrographic type 3.00 (Chwichiya 002, NWA 11750 and NWA 12957), the C3 chondrite El Médano (EM) 200, and metamorphosed ungrouped chondrites with petrographic types ranging from type 3 to type 5/6 (Los Vientos (LoV) 051, Sahara 00177, Coolidge, Mulga (West) by increasing order of thermal metamorphism). The studied "grouped" meteorites were either from rare groups, with no parent bodies associated yet (i.e., CH or Kakangari groups), or showed unusual petrographic features making them anomalous in their group. They were also helpful to verify that the method used in this paper gives results comparable to previous studies (Burbine et al., 2003; Dunn et al., 2010; Moskovitz et al., 2010; Vernazza and Beck 2016; Vernazza et al., 2017; Lucas et al., 2019; Takir et al., 2019). We selected five grouped carbonaceous chondrites: two CM chondrites, Aydar 003 (CM1/2) and NWA 11086 (CM-an), one CH3, Los Vientos (LoV) 200, and one CR3, NWA 12474. We also studied the Kakangari meteorite, the only fall in the small K group, as well as two Rumuruti chondrites: Awsserd (R4) and NWA 12472 (R3). We also analyzed the ordinary chondrite NWA 12334 (LL6-an). Finally, two primitive achondrites, NWA 6592 (lodranite) and NWA 12480 (acapulcoite), were added to complete the variety of studied samples.

The surface of asteroids can be made of dust, regolith, or boulders. For better comparison to these blocky and/or dusty surfaces, we measured raw and powdered samples for each meteorite. For powdered samples, we used around 200 mg of material, dry ground in an agate mortar, giving a gain size < 100 μm. Polished sections were used as a substitute for raw samples in the case where samples were only available in that form. The comparison between the spectra of the different types of samples will be discussed in this paper. Most studied samples were analyzed using the three samples preparation (raw, powdered and polished samples).

## 2.2 Petrography

Polished sections of all meteorites were observed with a Leica DM2500P petrographic microscope. Observations and maps were made using reflected light on thick polished sections. Modal abundances of chondrites components (chondrules, matrix, opaques) were evaluated by point counting, with 95% confidence interval around the modal abundance computed after Howarth (1998). Chondrule sizes were measured by contouring chondrules outlines on optical maps using the *Photoshop* software, followed by image analysis using the *ImageJ* software.

Backscattered Secondary Electron (BSE) images were taken on all meteorites. Chemical compositions were evaluated by Energy Dispersive Spectrometry (EDS) analyses. Both BSE and EDS analyses were obtained with a Hitachi S3000-N SEM equipped with a Bruker X-ray EDS at the Centre Européen de Recherche et d'Enseignement en Géoscience de l'Environnement (CEREGE).

Petrophysical parameters (magnetic properties) were also used for a better characterization of the studied meteorites. Magnetic susceptibility was measured with a MFK1 or KLY2 instruments depending on sample size. Saturation magnetization was determined from hysteresis loops measured with a Princeton model 2900 Micromag vibrating sample magnetometer (see Gattacceca et al., 2014 for a comprehensive description). All magnetic measurements were also performed at CEREGE.

## 2.3 Reflectance Infrared spectroscopy

### 2.3.1 Instrumentation

The Infrared (IR) spectroscopy analyses were performed at the Institut de Planétologie et d'Astrophysique de Grenoble (IPAG) using the SHADOWS instrument (Potin et al., 2018). We measured reflectance spectra under nadir incidence and with an observation angle of 30°. The instrument was used in a standard mode, enabling a spot size of about 5.2 mm in diameter. All spectra were normalized to infragold™ and spectralon™. For each sample, we measured the reflectance spectra in a range from 0.3 µm to 2.6 µm, with a step of 20 nm.

### 2.3.2 Terrestrial weathering

Terrestrial weathering of the meteorites can also influence the spectra and thus the comparison of meteorite and asteroid spectra. Typical terrestrial weathering products, iron oxides/hydroxides, generally exhibit absorption edges around 0.55-0.60 µm, with absorption bands at 0.5 and 0.9 µm and weaker absorption associated to OH- observable at 1.4 and 1.9 µm (Hiroi et al., 1993; Cloutis et al., 2010). As this weathering is only terrestrial, those products should be removed to improve the comparison of meteorite and asteroids infrared spectra. To reduce the effect of terrestrial weathering on meteorite spectra and ease the comparison with asteroid spectra, strongly weathered samples were treated with a leaching process (Cornish and Doyle, 1984; Lucas et al., 2019). Lucas et al. (2019) showed that EATG leaching treatment did not affect the silicates bands positions, and slightly increases the depths of the bands on the reflection spectra on anhydrous meteorites and pyroxene and olivine control samples. In this paper, we tested different leaching processes in order to choose the most efficient and least aggressive on carbonaceous chondrite samples (HCl, EATG and HCl+ EATG). We measured spectra from 0.35 to 4 µm to observe the effect on silicates as well as on hydrated minerals bands. We compared the spectra without leaching to the spectra for the three types of leaching. We observed the same silicates bands positions with a small increase in the depth of the silicate's bands. We also observed that the 3 µm band of hydrated minerals was not significantly modified by the leaching. Thus, we chose to follow the same methodology as Lucas et al. (2019) as it did not show any strong effect on the studied hydrated chondrites except for the removal of terrestrial alteration. We used around 200 mg or powdered samples left in 5 mL Ethanolamine Thioglycolate (EATG) 67% in Thioglycolate acid (TGS) 40% for 2 hours. After two to three EATG cycles for the most weathered samples, the samples were thoroughly rinsed with water first, then pure IPA before being dried.

#### 2.3.3 Separation of asteroids complexes and types

The comparison of asteroid spectra to meteorite spectra is limited by the resemblance of some asteroid complexes in reflectance spectra in the studied wavelength range. Especially, some meteorites, such as Chwichiya 002 or NWA 11750, could be linked to both X- and C-complex asteroids. Indeed, the types Cg, Cgh, Ch, Xk, Xc and Xe are not easily separated on component space methodology used by DeMeo et al. (2009), because their spectral features are weak. The study made in this paper is made at the first order, with spectral shape and parameters comparison, so the differentiation of the match of the two complexes of asteroids is not possible for all meteorites.

In greater details, some asteroid spectra of the same complex were too similar to be differentiated during the process of matching the spectra of meteorites to asteroids. Indeed, for some asteroids, the difference between two asteroid spectra is smaller than the difference to the closest matching meteorite. For example, Xe- or Xk-type asteroid spectra are very similar, so they will often be matched to the same meteorite. DeMeo et al. (2009) show that Xe- and Xk- type spectra are very similar and need the detailed study of a slight feature around 0.9-1 μm to differentiate the two types. In our study, we did not try to differentiate the two spectra, so they are both matched with the same meteorites.

Another limitation of the comparison of asteroids to meteorites is the complexity of the mineral composition of asteroids surfaces. An asteroid could be composed of different chondrite types, thus the spectra obtained from this asteroid corresponds to the average of the different type of mineralogy composing its surface. It could therefore be an average spectrum of different meteorite spectra. In consequence, the comparison of one asteroid spectrum to one meteorite spectrum is limited, as both scales are very different. For example, laboratory measurements of hydrated carbonaceous chondrites show band depths stronger than the associated Cg- and Cgh-type asteroids. If we rule out space weathering effects, it could be explained by the surface of the asteroids being made of both hydrated and non-hydrated carbonaceous chondrites (Beck et al., 2018).

#### 2.3.4 Space weathering

It is also discussed that optical properties of the surface of an asteroid can change with processes such as micrometeorite impact and sputtering due to solar wind (e.g., Chapman, 1996; Sasaki et al., 2001; Hapke, 2001). Those processes can influence the spectra of the asteroids and create differences between asteroid and meteorite spectra. Indeed, this process is understood between S-type asteroids and their linked ordinary chondrites (Sasaki et al., 2001, Strazzulla et al., 2005, Marchi et al., 2005, Brunetto et al., 2006, Vernazza et al., 2009a). It could therefore also influence the matching of ungrouped meteorites with other asteroids types. In the case of silicate-rich asteroids, space weathering reduces silicates bands depths, and increases the slopes of the VIS-NIR spectra. In the case of carbonaceous asteroids, space-weathering effects are likely present, and would have similar effect, but of lower magnitude (Lantz et al., 2017; Rubino et al., 2020). We also believe that space weathering is stronger on the asteroids than on the meteorite samples studied on earth. Thus, the asteroid spectra could have a redder VIS-NIR slope and show shallower absorption bands than the associated meteorite. Since this effect is hardly quantifiable on the asteroid's spectra, to ease the comparison of meteorites and asteroids, we chose to make the hypothesis that asteroids have low space weathering.

### 2.4 Spectral parameters and establishment of methodology

To analyze the reflectance spectra, a set of well-defined spectral features was determined to allow for meteorite-meteorite and meteorite-asteroid comparisons. This was done as described in Eschrig et al. (2020) which follows earlier works by Cloutis et al. (2012). The spectral features include several spectral band depths, positions and slopes which were consistently determined using a python script. The 1 μm and 2 μm band ranges were fit with baselines by determining the maxima on either side of the bands. Subsequently, the 1 μm and 2 μm slopes ($nm^{-1}$), band depths (%) and positions (nm) were determined. The slopes were taken from the 1 μm and 2 μm band baseline slopes and the band depths and positions were determined following Clark et al. (1999). In case of spectra with particular shapes, the positions of the margins of the baseline were adapted manually as seen fit by the author, using spectra shapes, silicates bands positions and slopes changes.

## 3 Petrography

In order to introduce some logic in the population of studied meteorites, we grouped them into six meaningful petrographic groups, labelled A to F in the following (Table 1). The key petrographic features are summarized in Table 1, and BSE images are provided in Figures 1 to 5. Two magnetic parameters (saturation magnetization, $M_S$, and magnetic susceptibility, $\log\chi$, with $\chi$ in $10^{-9}$ $m^3/kg$), which are quantitative proxies to the abundance of ferromagnetic minerals (mostly metal and/or magnetite in the studied meteorites), are also given in Table 1. It is noteworthy that these are two intrinsic magnetic parameters that cannot be affected by exposure to magnetic fields (such as magnets). However, for metal-bearing meteorites, they decrease with increasing terrestrial weathering due to the progressive destruction of metal (e.g., Rochette et al., 2003). The petrographic group A (Fig. 1) comprises ungrouped carbonaceous chondrites that do not show evidence of significant thermal metamorphism nor aqueous alteration: Chwichiya 002, NWA 11750, NWA 12957, and Acfer 094. Chwichiya 002, NWA 11750, and NWA 12957 show roughly similar petrography indicative of very low or absent aqueous alteration and absence of significant thermal metamorphism, based on Raman spectroscopy, X-Ray Diffraction (XRD) and Infrared (IR) transmission spectroscopy, thus their petrographic type 3.00. They show broadly CM-like petrography, with high matrix abundance (63.3-74.0 vol%), low chondrule abundance (12.9-26.0 vol%), and small chondrule apparent diameter (average in the range 240-480 µm) (Table 1), but they differ from CM chondrites in their oxygen isotopic compositions and no evidence for significant aqueous alteration (Gattacceca et al., 2020b). Acfer 094 is classified as C2-ung, with trace element and petrography resembling CM2 chondrites. However, it is pooled here with C3.00 chondrites because XRD does not evidence a significant amount of phyllosilicates, and Acfer 094 contains more pre-solar SiC grains (very sensitive to aqueous alteration and thermal metamorphism) than any other meteorite, suggesting minimal thermal or aqueous processing (Newton et al, 1995). It shows a slightly different silicate composition, with addition of plagioclase compared to the other 3 meteorites from this group A, and a slightly lower matrix abundance (60.9 vol%).

The petrographic group B (Fig. 2) is composed of CM-like meteorites, similar to the group A, but with meteorites showing significant traces of aqueous alteration. This group comprises Aydar 003 (CM1/2), NWA 11086 (CM-an), El Médano (EM) 100 (C2-ung), NWA 8781 (C-ung), NWA 5958 (C2-ung), and EM 200 (C3). Aydar 003 (CM1/2) shows small chondrules, mineral fragments, Calcium–aluminum-rich inclusion (CAIs), sulfides, and magnetite in an abundant fine-grained phyllosilicate-rich matrix, like CM chondrites (Gattacceca et al., 2020c). NWA 11086 (CM-an) also has a petrography resembling CM2 chondrites. It shows no phyllosilicates by XRD, but VIS-NIR reflectance spectra (see below) reveal the presence of a low amounts of phyllosilicates (Gattacceca et al., 2019). NWA 5958 (C2-ung) has affinities with CM2 chondrites but with oxygen isotopes composition plotting on an extension of the carbonaceous chondrites anhydrous minerals mixing line (CCAM) (Jacquet et al., 2016). EM 100 (C2-ung) also resembles CM2 chondrites but has unusual oxygen isotopic compositions (Ruzicka et al., 2015). EM 200 (C3 with affinities with CO3) shows no phyllosilicates by XRD, but small amounts of aqueous alteration on IR spectra of the matrix (Ruzicka et al., 2015). NWA 8781 (C-ung) is a peculiar, it shows small chondrules, absence of magnetite, unusual oxygen composition and absence of CAI. Phyllosilicate are not found in the matrix, and ferroan olivine have 0.2-0.4 wt% $Cr_2O_3$ (Gattacceca et al., 2017) indicating a subtype 3.0 (Brearley and Grossman, 2005).

The petrographic group C (Fig. 3) comprises ungrouped chondrites that show traces of aqueous alteration, as previous group B, with the addition of small thermal metamorphism traces: NWA 12474 (CR3.1), Dho 2066 (C-ung), Kakangari (K); and chondrites with no matrix, LoV 200 (CH3), and Sierra Gorda 009 (Chondrite-ung). NWA 12474 shows well-defined large chondrules with composite opaque grains in an abundant fine-grained matrix, showing very slight metamorphism and no phyllosilicates, based on Raman spectroscopy and XRD (Gattacceca et al., 2020c). Kakangari has sharply defined chondrules surrounded by a high proportion of unequilibrated matrix dominated by magnesian pyroxene (enstatite) and magnesian olivine, with few chondrules (Weisberg et al., 1996; Barosh et al., 2020). It experienced secondary aqueous alteration and metamorphism on its parent body (Barosh et al., 2020). The carbonaceous chondrite Dho 2066 has been classified as C-ung but has been recently associated with the elusive CY group (Ikeda, 1992; King et al., 2019). It has unusual oxygen isotopic composition and mineralogy (absence of orthopyroxene, atypical olivine compositions compared to CMs), and low metal abundance (<1 vol%). It is probably paired with Dho 735, texturally resembling dehydrated CM2 chondrites, but with different bulk chemistry and oxygen isotopic composition than CM2 chondrites (Ivanova et al., 2010). LoV 200 (CH3) shows small packed chondrules, with almost no matrix, and abundant metal.

Sierra Gorda 009 is somewhat intermediate between ordinary and enstatite chondrites. It resembles "Low-FeO" or "G-chondrites" (Weisberg et al., 2015, Ivanova et al., 2020), with very high content of metal, large pyroxene-rich chondrules and no fine-grained matrix. Its silicates are more reduced than in H chondrites, but less reduced than in enstatite chondrites.

The petrographic group D is formed by two members of the relatively rare Rumuruti chondrites group: Awsserd (R4) and NWA 12472 (R3) (Fig. 4a and Fig. 4b). Rumuruti chondrites show small chondrules in an abundant matrix, with Fe-rich olivine as the main silicate and sulfides as the main opaque phase (Fig. 4a and 4b). Both studied Rumuruti chondrites are significantly metamorphosed: Awsserd has a petrographic type 4, and NWA 12472, based on the Cr2O3 content in the olivine of 0.13±0.05 wt.% (Gattacceca et al., 2020c) has an estimated subtype ≥3.2 (Brearley and Grossman, 2005).

The petrographic group E gathers significantly metamorphosed ungrouped chondrites (Fig. 4c to 4f). It contains Sahara 00177 (C3/4-ung), LoV 051 (C3-ung), Coolidge (C4-ung) and Jah 846 (OC3). Sah 00177, Coolidge and LoV 051 have relatively similar textures showing broad similarities with the texture of reduced CV and CR chondrites (McSween and Richardson, 1977; Ruzicka et al., 2017). However, they have higher metal/magnetite ratio than CVred chondrites, lower matrix abundance than CVred and CR chondrites, and smaller chondrules than CVred and CR chondrites (Metzler et al., 2020). JaH 846 has chondrules dominated by orthopyroxene and silica polymorphs (Table 1) and an oxygen isotopic composition that is clearly separated from the composition of the three groups of ordinary chondrites. The $Cr_2O_3$ content of its ferroan olivine (0.08±0.06 wt%, n=8) indicate a petrographic type ≥3.2 (Gattacceca et al., 2020b)

Finally, petrographic group F gathers highly metamorphosed meteorites: NWA 12334 (LL6-an), Mulga West (C5/6-ung), NWA 12480 (acapulcoite) and NWA 6592 (lodranite) (Fig. 5). The metamorphosed anomalous ordinary chondrite NWA 12334 is an anomalous member of the LL group. Compared to other LL chondrites, it has unusually high Fa content in olivine ($Fa_{34.1\pm0.3}$), low magnetic susceptibility, and unusual opaque mineralogy featuring magnetite and pyrrhotite instead of Fe,Ni metal (Gattacceca et al., 2020c). Mulga West shows a significantly recrystallized CV-like petrography, with resorbed chondrules in an abundant fine granoblastic silicate matrix. NWA 12480 shows a strongly recrystallized texture composed of mainly orthopyroxene (typical size 400 μm), Ca-pyroxene, plagioclase (typical size 200 μm) displaying triple junctions (McSween, 1977). Finally, NWA 6592 is composed of a coarse-grained of 1-2 mm mineral aggregate of olivine, pyroxene, feldspar, and opaques, with texture showing triple junctions (Keil et al., 2017).

# 4 Infrared spectroscopy

## 4.1 Infrared spectra of powdered samples

In this section we present the different spectra obtained by reflectance measurement on meteorite powdered samples (Fig. 6). Spectral groups were made based on spectral similarities, in order to simplify the comparison to asteroid spectra and interpretations (Table 2). In addition, to ease the comparison of meteorite to asteroid spectra, we use the equivalent geometric albedo (Fig. 7). The equivalent geometric albedo of the meteorites is obtained by using the reflectance value measured under standard conditions (at 550 nm, a phase angle g=30°) and multiplying by a factor from Figure 3 of Beck et al. (2020), having the corresponding reflectance at very low angle (g<2°). The main part of the petrographic group (Fig. 1-5) mirrors the spectral grouping, except for a few meteorites (Acfer 094, NWA 8781, JaH 846, and NWA 12334), showing that spectra features reflect mineral abundances and compositions.

The first spectral group is composed of spectra showing no absorption bands, or very broad and shallow absorptions around 1 µm only (Fig. 6a). They show a lack of 2 µm absorption bands, and a flat spectral VIS-NIR slope. They also show low albedo at 550 nm typically ≤ 0.1, except for EM 200 with an albedo slightly higher around 0.14 (Fig. 7). They all show a reflectance maximum around 600 nm, with a small positive slope from 1 µm to higher wavelengths. The meteorites in this spectral group are the NWA 11750, NWA 12957 and Chwichiya 002, all classified as C3.00-ung. They correspond to the petrographic group A (Fig. 1), without Acfer 094, and with the addition of EM 200 of petrographic group B.

We interpret these featureless spectra to reflect the high abundance of matrix (63.0, 73.4 and 74.0 vol%) rich in phases that show no spectral features (carbonaceous matter, opaques) (Table 1). The lack of spectral features can also be explained by the low abundance of chondrules (12.9, 24.8 and 26.0 vol %), with small to medium apparent average size (240-480 µm, average 340 µm), rich in silicates, that would have deep absorption bands around 1 µm and 2 µm (Table 1). The overall darkness can be explained by the high abundance of matrix that hides silicate absorption bands. The lack of phyllosilicate signature agrees with the classification as type 3.00 that denotes the absence of significant hydration of the meteorites of the petrographic A. EM 200 has no phyllosilicates signatures on XRD, but very low amounts of phyllosilicates on IR transmission of its matrix. Its proportions of phyllosilicates is thus very low and does not show on the reflectance spectra, making it similar to the 3.00 chondrites.

The second spectral group is composed of NWA 5958, Aydar 003, EM 100, NWA 11086. These meteorites are all in the petrographic group B. They show a rather sharp maximum around 0.6 µm, together with a shallow absorption band around 1 µm (Fig. 6b). While there is variability in spectral VIS-NIR slopes, these four meteorites show spectral signatures of phyllosilicates at 0.7, 0.9 and 1.1 µm (absorptions by $Fe^{2\pm}Fe^{3+}$). The case of EM 100 is particular since signatures of phyllosilicates can be also found at 1.9 µm (absorptions by $H_2O$) and 2.3 µm (absorptions by X-OH). Those spectra show small absorptions bands, which can be explained by a high proportion of matrix (44.1-75.5 vol%, average 55.8 vol%). The phyllosilicate spectral signatures in this spectral subgroup are in agreement with the significant aqueous alteration of the matrix of this petrographic subgroup. The equivalent albedo (at 550 nm) of EM 100 and NWA 11086 are relatively high compared to C2 chondrites and meteorites from spectral group 1 (Fig. 7), which may be related to a lower abundance of opaque phases (7 vol% for EM 100 and 2 vol% for NWA 11086).

The third spectral group has higher VIS-NIR spectral slope values than the first two groups, and deeper absorption band around 0.9-1 µm, as well as slight to no absorption bands around 2 µm (Fig. 6c). They also show low albedo at 550 nm, slightly higher than the first spectral group, ranging 0.07-0.12, with Sierra Gorda 009 as an exception (0.19) (Fig. 7). The meteorites in this spectral group are most of the meteorites of the petrographic group C: Dho 2066, LoV 200, Sierra Gorda 009, NWA 12474, and the rare chondrite Kakangari (Fig. 3); with the addition of Acfer 094 from petrographic group A (Fig. 1) and NWA 8781 from petrographic group B (Fig. 2).

Compared to the two first spectral groups, showing very small absorptions features, this spectral subgroup has higher VIS-NIR slopes and deeper absorption bands around 0.9 µm and small to no absorption bands around 2 µm, which agrees lower abundances of matrix (0-66.8 vol%, average 41.8 vol%) and higher chondrule abundances (22.7-74.5 vol%, average 41.5 vol%) and size (130-1050 µm, average 485 µm) (Table 1). Indeed, with a general higher proportion of silicates and less fine-grained matrix, spectral features around 0.9 µm and 2 µm appear deeper on the spectra. In greater detail, Kakangari, Dhofar 2066 and NWA 12474 have spectra reflecting a high matrix abundance, hiding the silicates absorption bands. Kakangari has also a matrix composition rich in enstatite, a magnesium-rich pyroxene with less absorption bands around 2 µm than iron-rich pyroxene. LoV 200 show a positive VIS-NIR slope with shallow and broad features near 950 nm. Its spectrum reflects the absence of matrix with abundant (66 vol%) chondrules, composed of a high proportion of pyroxene in form of enstatite. In addition, the abundant opaques (24.4 vol%), mostly under the form of metal, have no IR absorption signature but can explain the overall red VIS-NIR slope. Acfer 094 and Sierra Gorda 009 have similar flat to slightly negative VIS-NIR slopes with a small absorption band around 0.9 µm. Sierra Gorda 009 has a spectrum reflecting the absence of matrix with abundant chondrules creating the absorption around 900 nm. The lack of broad absorption band around 2 µm is explained by the pyroxene composition low in iron ($Fe_{1.42\pm0.39}Wo_{0.88\pm0.63}$) that have shallower absorption bands that iron-rich pyroxene. We can see a small absorption band around 1.9 µm due to potential iron oxides left from terrestrial weathering even after the leaching process. Acfer 094 is part of the petrographic group A but spectral group 3, due to its more positive slope and deeper absorption band around 0.9 µm than the three other meteorites of petrographic group A. Indeed, Acfer 094 is spectrally different than petrographic group A, explained by its different silicate composition, richer in iron, in addition to less abundant matrix and more abundant chondrules, creating deeper absorption bands. NWA 8781 is from petrographic group B, but has spectral characteristics from spectral group 3. It has a slightly deeper absorption band and more positive VIS-NIR slope than other meteorites of the petrographic group B. This can be explained by a lower matrix abundance, with higher proportion of silicates than other meteorites of petrographic group B.

The spectral group number 4 is defined by the spectra showing narrow absorption bands around 1050 nm with sharp walls of the absorption band and usually slightly positive general VIS-NIR slopes (Fig. 6d). They have medium equivalent albedo at 550 nm, from 0.1 to 0.18 (Fig. 7). The meteorites in this spectral group are the Rumuruti chondrites, Awsserd and NWA 12472 from petrographic group D. Awsserd and NWA 12472 display very similar spectra showing a 1050 nm absorption band and a broad and shallow absorption band around 2000 nm, with positive VIS-NIR slopes. The fourth spectral group spectra are explained by the iron-rich olivine dominated chondrules ($Fa_{37.2}$ and $Fa_{40.9}$) (Table 1). Indeed, Rumuruti chondrites have the most iron-rich olivines in all chondrites, creating deep absorption bands around 1 μm.

The fifth spectral group shows spectra with strong and relatively narrow absorption band around 1 μm with a faint absorption around 1900 nm (Fig. 6e). The visible maximum has a rounded shape around 750 nm and the general VIS-NIR slope or the spectra can vary from slightly negative to slightly positive but all of them show similar absorption bands. NWA 12334, LoV 051 and Coolidge have low abundance of matrix (0, 17.2, 19.3 vol%) (Table 1), creating a spectrum dominated by silicates with deep absorption bands around 1 and 2 μm. The smaller chondrules of Coolidge explains the less pronounced bands of its spectrum compared to LoV 051. NWA 12334 is from petrographic group F (Fig. 5) but has spectral similarities from spectral group 5 (Fig. 6e). It has indeed shallower absorption bands than the other meteorites from petrographic group F (Fig. 6f). The equivalent albedo is high for NWA 12334 (0.16), LoV 051 (0.15) and Coolidge (0.14) (Fig. 7).

The sixth spectral group (Fig. 6e) comprises the meteorites from petrographic group F (Fig. 5), the primitive achondrites NWA 12480 (Acapulcoite) and NWA 6592 (Lodranite), and the ordinary chondrites JaH 846 from petrographic group E (Fig. 4f). This spectral group shows spectra with similar absorption bands positions as the previous spectral group, with deeper absorption and more positive VIS-NIR slopes (Fig. 6f). Indeed, those meteorites have recrystallized matrix due to metamorphism. Those recrystallized silicates explain the deep absorptions bands in this group. Members of this spectral group have variable equivalent albedo at 550 nm, with medium albedo for JaH 846 (0.12) and higher albedo for NWA 12480 (0.4) (Fig. 7). JaH 846's spectrum has a deep and narrow absorption band around 950 nm, with a broad absorption band around 1950 nm reflecting its unusual high proportion of pyroxene, with some chondrules dominated by pyroxene (Table 1). NWA 12480 shows the deepest absorption band around 950 nm, with a deep band around 1950 nm due to pyroxene-dominated composition as well.

## 4.2 The impact of sample preparation

The surface of most asteroids is covered with regolith (e.g. (132173) Ryugu, Jaumann et al., 2019), explaining why laboratory meteorite spectra are often done on powdered samples. However, asteroids can also present larger blocks on their surface, or even surfaces almost devoid of regolith such as asteroid (101955) Bennu (Lauretta et al., 2019). Therefore, we chose to measure IR reflectance spectra of raw and polished section samples in addition to powdered samples. This allows assessing weather sample preparation changes the asteroid-meteorite spectral connections, an aspect that is rarely discussed in literature.

Some meteorites, achondrites or ordinary chondrites, have powdered samples with higher reflectance spectra than the raw and polished section samples (Fig. 8A), but other samples, especially carbonaceous chondrites, show higher reflectance on raw sample spectra compared to powdered samples (Fig. 8B). In addition, independently from the chondrites group, the powdered samples show a redder spectrum than the raw and polished sections (Fig. 8, Fig. 9). Indeed, powdered samples have higher slopes at 1 μm as well as 2 μm than polished sections and raw samples for all spectral groups (Fig. 9). We know that grain size, more generally texture, impact the VIS-NIR slope and the depth of absorption bands of a spectrum (Ross et al., 1969; Johnson and Fanale, 1973, Cloutis et al., 2018). The redder VIS-NIR slope of powdered samples compared to raw samples was observed by Beck et al. (2021) on the CM chondrite Aguas Zarcas. This redder VIS-NIR slope of powdered sample spectra can be explained by red sloped phases becoming more visible when finely mixed with other minerals (Ross et al., 1969) (Fig. 9). It could also be due to the grinding changing the mixture and the grain sizes. Indeed, when grinding the sample, the powder becomes an intimate amalgam, linear in a single-scattering albedo space. Raw samples, on the contrary, give a geographical linear in reflectance space. Thus, the grinding can make the matrix of the meteorites more visible than on raw samples, darkening and reddening the spectra. The organic matter and the alteration phases in their matrix being more visible on powdered samples, it will darken and redden the spectra and hide the silicates absorption bands. This explains why most carbonaceous chondrites (spectral group 1-3), with high matrix abundances, have darker and redder spectra for powdered samples than raw samples (Fig. 7). Additionally, polished sections are one of the most used preparations of meteorites as it is the easiest way to assess the petrographic context. Also, polished sections are easily available from meteorite repositories, whereas raw samples for rare and small meteorites may be more difficult to obtain in loan. Thus, being able to measure infrared reflectance spectra of a polished section is undoubtedly interesting. For polished sections, sample preparation influences the reflectance spectra for carbonaceous chondrites, especially spectral group 1 to 3, but less for other chondrites. Indeed, for carbonaceous samples, it creates redder VIS-NIR slopes than raw samples (Fig. 8B, Fig. 9). Due to the influence of polished section preparation on the reflectance spectra measurement, if raw or powdered sample present a different comparison than the spectra on polished sections for the same meteorite, we give more weight to the raw and powdered samples spectra.

## 4.3 Comparison of ungrouped and grouped meteorites

The samples chosen in this study are ungrouped chondrites or meteorites from rare groups. In addition to their petrographic and chemical specificities, we now compare their spectral properties with that of well-studied meteorite groups to determine if their particularities stand out in infrared spectroscopy.

For the first spectral group, NWA 12957, NWA 11750 and Chwichiya 002 have broadly similar petrography but differ by oxygen isotopic composition and opaque abundances (as reflected also by their different magnetic properties) (Fig. 1, Table 1). Thus, their spectra are similar, except for a small difference in position of the maximum peak around 600 nm. The absence of clear absorption bands makes their spectra similar to typical spectra of CM, CO, CV or CR chondrites, but their spectra do not resemble any of those chondrites' spectra (Fig. 10a). For the second spectral group, the closest meteorite spectra are of CM chondrites (Fig. 10b). We can see that indeed Aydar 003 and NWA 5958 spectra are similar to a CM spectrum, only with bluer VIS-NIR slopes. NWA 5958 is known to have CM-related petrography and bulk chemistry (Jacquet et al., 2016), so the IR spectra are similar. A difference between NWA 5958 and CM chondrites is the absence of small absorption band at 750 nm and the bluer VIS-NIR slope in the former. NWA 11086, classified as CM-an, has a spectrum resembling CM chondrites, with a slightly more positive VIS-NIR slope. Finally, EM 100 has a spectrum resembling CM chondrites, but with a very strong VIS-NIR slope and appears overall much brighter than typical CM chondrites.

In the third spectral group, LoV 200 has a spectrum resembling enstatite chondrites (EC) spectra, similar to another CH chondrite studied (Cloutis et al., 2012) (Fig. 10c). Indeed, LoV 200 has a high proportion of pyroxene in the form of enstatite, which is a common feature with enstatite chondrites. NWA 12474 has an IR spectrum similar to CR2 chondrites, with a bluer VIS-NIR slope (Fig. 10c). Kakangari has a spectrum with features in between those of ordinary chondrites and enstatite chondrites (Fig. 10c). This is explained by the high matrix abundance resembling carbonaceous chondrites hiding the absorption bands, and with a matrix rich in enstatite, resembling EC (Table 1). It has a metal abundance closer to ordinary chondrites, in addition to olivine and pyroxene compositions in between EC and H ordinary chondrites (Weisberg et al., 1996) (Table 1). Acfer 094 has a spectrum resembling the Kakangari spectrum, with a faint absorption around 0.9 µm, and resembles spectra of CR chondrites (Fig. 10c). Dho 2066 has a petrography resembling CM chondrites, but shows shallower absorption around 1 µm and a more positive VIS-NIR slope, resembling CH meteorites or CY meteorites (King et al., 2019) (Fig. 10c). NWA 8781 resembles most CR meteorite spectra, with shallower absorption bands that could be explained by the lower abundance of silicates compared to CR chondrites, or a low thermal metamorphic grade (Eschrig et al., 2020) (Fig. 10c). Sierra Gorda 009 has a spectrum similar to L chondrites (Fig. 10c). It has indeed no matrix and abundant chondrules, explaining the spectrum resembling ordinary chondrites (Table 1).

For the fifth spectral group, LoV 051 spectrum resembles ordinary L meteorites (Fig. 10d). It has low matrix abundance, with high proportion and larger chondrules than CR chondrites (Table 1). NWA 12334 is a LL-an and its spectrum resembles LL chondrites spectra well (Fig. 10d).

Finally, for the sixth spectral group, JaH 846 and NWA 12480 have deep absorption bands around 1 µm and 2 µm, dominated by silicates, and are thus closer to ordinary chondrite spectra (here compared to L or LL spectra) (Fig. 10e).

# 5 Potential parent bodies

## 5.1 Qualitative comparison of reflectance spectra of meteorite powders to asteroid observations

In this section we present a qualitative comparison of infrared spectra of the studied ungrouped or rare meteorites to asteroids, in order to find resemblances, and hence potential parent bodies. For this comparison we used DeMeo et al. (2009) taxonomy based on more than 400 infrared spectra of asteroids in the 0.45 to 2.45 µm range.

As surfaces of large asteroids are usually covered by regolith, and usually parent body research is done on powdered samples, we compare, in a first step, asteroid reflectance spectra with reflectance spectra of our meteorite powders. We first match the spectra qualitatively (qualitative match, called QM in the following), and then compare quantitative spectral parameters (§5.2). The preferred asteroid matches to the meteorite spectra are presented for all samples in Table 2. In addition, to complete the comparison of meteorites to the asteroids, we use the equivalent geometric albedo and test if different spectral groups of meteorites match the albedo of the associated asteroid groups (Fig. 7). This low angle reflectance is then comparable to the albedo of the asteroids.

For the first spectral group, we observe that the powdered meteorite reflectance spectra have a QM to C-complex asteroids (Table 2, Fig 6a). C-type reflectance spectra have low to medium VIS-NIR slope with small to no features (DeMeo et al., 2009), consistent with carbonaceous chondrites. In details, Ch- and Cgh-type asteroids are classically linked to CM chondrites (Vilas and Gaffey, 1989; Vilas et al., 1993; Rivkin, 2012; Rivkin et al., 2015; Lantz et al., 2013; Burbine, 2014; McAdam et al., 2015; Vernazza et al., 2016; Vernazza et al., 2017), but they have also been matched with multiple other meteorites groups, such as ureilite (Jenniskens et al., 2009), CI (Johnson and Fanale, 1973; Cloutis et al., 2011), CR (Hiroi et al., 1996; Sato et al., 1997), K (Gaffey, 1980). In contrast, Cg- and B-type asteroids have not been matched to any meteorite group yet (Vernazza et al., 2015). This first spectral group matches best to C- and Cg-types. NWA 12957 is closer to Cg-type with a maximum around 650 nm and a pronounced UV drop-off (DeMeo et al., 2009). Chwichiya 002, EM 200 and NWA 11750 match C-type asteroid spectra. In addition, the spectra of the first spectral group of meteorites show low equivalent albedo, similar to low albedo asteroids C-, B- and D-types (Fig. 7), consistent with the match to C-complex (Fig. 6).

The second spectral group also has a QM with C-complex asteroids (Fig. 6b). Aydar 003 shows a good resemblance with Ch-type asteroids, with the same slight positive VIS-NIR slope and the small absorption band around 750 nm. NWA 5958 powder spectrum has no clear match, in between B-type and Ch-type asteroids. NWA 11086 has a slight absorption bands around 1 μm and no features around 2 μm, resembling C-complex spectra, but its positive VIS-NIR slope does not resemble the C-complex. We were therefore unable to propose a good match for this sample. Similarly, EM 100 cannot be matched to any asteroid. Indeed, its spectrum resembles Ch-type asteroids but with a much higher spectral VIS-NIR slope, especially visible on powdered samples. The second spectral group has a general low equivalent albedo (Fig. 7), similar to asteroids of the C-, B- and D-complexes. NWA 11086 and EM 100 are exceptions and show higher albedo (Fig. 7), potentially due to very positive VIS-NIR slopes.

The third spectral group is composed of meteorites with a QM with the X-complex asteroids (Fig. 6c). X-complex asteroids have spectra with medium to high VIS-NIR slope. Xc-type asteroids present featureless spectra, and are linked to enstatite chondrites and aubrites (Zellner 1975, Zellner et al., 1977, Vernazza et al., 2009b, 2011, Shepard et al., 2015; Vernazza et al., 2016). Xk- and Xe-type asteroids show features around 0.9-1 μm, associated to low-FeO pyroxene (Clark et al., 2004; Gaffey et al., 1989; DeMeo et al., 2009). Xk-type asteroids are supposed parent bodies of mesosiderites, but Xe-type have no associated meteorite groups yet (Vernazza et al., 2009). LoV 200 matches Xe- or Xk-types asteroids, only with a flatter VIS-NIR slope for the raw sample. Dho 2066 spectrum is very similar to LoV 200 and resemble Xe-, Xk- or Xc-types. Acfer 094, NWA 8781, Sierra Gorda 009 and Kakangari resemble the Xe- and Xk-types spectra but with slight differences in band positions and depths. K meteorites were associated to C-complex by Gaffey (1980), but here we rather associate the Kakangari sample to X-type complex as the 1 µm band position resembles more X-complex asteroids. X-complex asteroids are not represented in Fig. 7. Their albedo is divided into three classes with distinct albedo: E (> 0.30) M (0.0075<M< 0.30) and P (<0.075) (DeMeo and Carry, 2013). The third spectral group of meteorites has an equivalent geometric albedo ranging between 0.07-0.2, associated to M-type. The spectral matching to X-type asteroids is consistent with the albedo comparison, more specifically to the M-subcategory.

The meteorites of the fourth spectral group have a QM with the K-type asteroids (Fig. 6d). This type is the potential parent body of CV, CO, CR and CK meteorites (Cruikshank and Hartmann, 1984; Sunshine et al., 2007; Clark et al., 2009; Cloutis et al., 2012; Vernazza et al., 2016). In this study, the meteorites linked to the K-type asteroids are R chondrites. Sunshine et al. (2007) linked Rumuruti meteorites to the A-type asteroids using band positions, chemical composition of olivine and melting models. In this study, we found that the A-type reflectance spectrum shows too deep absorption bands and too high VIS-NIR slope compared to these two samples. The two meteorites have absorption bands around 1050 nm that is attributed to olivine. The equivalent geometric albedo of those Rumuruti are in the same range of the K-type asteroids (and also L-type asteroids), which corroborate their match. This could be due to a difference in grain size. A-type asteroids could have larger grain sizes, thus deeper absorption and brighter albedo, compared to K-type and powdered samples or Rumuruti with smaller grain sizes.

The fifth spectral group of meteorites has a QM with the S-complex (Fig. 6e). S-complex asteroids are the most common asteroids in the inner region of the asteroids belt (between 2 and 2,5 AU of semi-major axis) (Gradie and Tedesco 1982; DeMeo and Carry 2013). This complex is usually associated to ordinary chondrites (Vernazza et al., 2016). It has also been associated to mesosiderites, acapulcoite/lodranite and winonaite (Gaffey et al., 1993), or to angrite (Rivkin et al., 2007). Differentiating between the asteroid types inside the S-complex (such as S, Sv, Sr, Sq, Sw) is difficult without using multiples parameters in a PCA (DeMeo et al., 2016). In this study, some spectral groups were not differentiated. We only made a difference between the S-, Sr-, and Sv-types on one side, and the Sa- and Sq-types on the other side. Those two latter show broader absorption bands around 1 µm, and Sa-type asteroids show a shallower absorption around 2 µm. LoV 051 is a good match to S- and Sr-type asteroids with less positive VIS-NIR slopes. Its equivalent geometric albedo resembles lower albedo S-type asteroids (Fig. 7). Coolidge has a similar spectrum as LoV 051 with less absorption bands, thus not match to any clear asteroids, but is closer to S-type asteroids. NWA 12334 also matches the S- and Sr-types spectra on powdered sample. It shows a spectrum similar to ordinary chondrite spectra, but the type 6 of the chondrite creates slightly deeper bands.

Finally, the sixth spectral group of meteorites is a QM to S-types with deeper absorption bands, such as Sa-, Sq- and Sv-types (Fig. 7f). Sv-type asteroids have spectra intermediate to S- and V-types. They have the narrowest features of the S-complex, resembling V-type spectra. They are pyroxene-rich spectra showing deep absorption bands around 1 and 2 µm. This feature is a good match with the spectra of NWA 12480. JaH 846 matches the Sq- and Sv-types, with deeper absorption bands. The sixth meteorite spectral group's equivalent geometric albedos are variable. NWA 12480 has spectra similar to Sv- and V-types asteroids. Its equivalent geometric albedo is high and is closer to V-type asteroids (Fig. 7). NWA 12334 has an equivalent geometric albedo close to lower albedo S-type asteroids corresponding to its spectra shape associated to S- and Sq-types asteroids (Fig. 7). Finally, the spectra of JaH 846 are similar to S-type asteroids. Its equivalent geometric albedo is slightly lower than the albedo of the S-complex (Fig. 7). Indeed, the albedo of S-type asteroids are higher than the albedo of type 3 ordinary chondrites, usually linked to S-complex. This cannot be explained by space weathering, as it darkens the surface of the asteroids and changes the VIS-NIR slope of the spectra (Pieters and Noble, 2016). An explanation is that the S-complex asteroids are not covered by type 3 ordinary chondrite material, but more thermally processed material of type 4 or above (Beck et al., 2020). This agrees with the hypothesis of a fast accretion followed by fragmentation and brecciation of the surface of S-type asteroids (Vernazza et al., 2014).

## 5.2 Quantitative comparison of IR parameters of meteorite powders and asteroid spectra

To improve the comparison between spectra from meteorites and asteroids, we compared quantitative spectra parameters. We used the following parameters: 1 μm and 2 μm bands positions (nm), the 1 and 2 μm bands depths (%), the VIS-NIR slope of the spectra in the range of 1 μm and 2 μm ($nm^{-1}$). Each spectral parameter was plotted against each other parameter to investigate every combination (e.g. Fig. 12).

We consider that a meteorite sample is matching an asteroid spectrum when the spectral parameters of the asteroid are in the range of the meteorite parameters (Fig. 11, Fig. 12). For the band positions, the range is limited by the resolution of asteroid spectra used in our work. The DeMeo et al. (2009) endmember spectra are acquired with a step of 50 nm in wavelength, while the resolution of the spectra of the meteorites is better, with a step of 20 nm. For the band depths and the slope parameters, the ranges are defined by standard deviations. The standard deviation is calculated for each parameter, using all 25 meteorites of the same sample preparation. Thus, the standard deviations of one parameter represent the possible variation of compositions in-between meteorites, without being influenced by the sample preparation (Fig. 12, Fig. 13). As a result, if the difference between the parameters of an asteroid and of a meteorite is smaller than the standard deviation of all meteorites, we considered that the asteroid and the meteorite have similar parameters, i.e. compositions. The range of possible matches for each meteorite describes an ellipsoid (semi-axis a, b) in each plot (Fig. 11, Fig. 12). A match between an asteroid (x, y) and a meteorite (h, k) is thus described by the equation : $\frac{(x-h)^2}{a^2} + \frac{(y-k)^2}{b^2} \leq 1$. The result of the parameters matching is reported in Table 2 and (Fig. S1).

We computed which asteroids have the highest recurrence match (HRM) to the studied meteorites (Table 2). We defined a matching score calculated as the number of plots where a certain asteroid spectrum matches a specific meteorite, normalized to the total of plots where the meteorite is compared, the latter depending on the number of features of the spectrum (Table 2). Indeed, not all spectra show features at 1 or/and 2 µm. When the band depths were < 2.5%, we considered that as no absorption bands, thus no feature. The first, second and third spectral group show no 2 µm band and some of them no 1 µm features (Fig. 6a, 6b and 6c). Spectra showing no absorption bands around 1 as well as around 2 µm can only be compared through slope parameters, lowering the matching score. The three first spectral groups of meteorites also display very shallow 1 µm band depths < 4.5 %, with no clear silicate absorption bands. Thus, we chose to compare those meteorite spectra only to asteroid spectra showing no clear silicates absorption bands: C- and X-complexes. The spectral group 4 to 6 show clear silicates absorption bands (Fig. 6d, 6e and 6f). They display 1 µm band depths (6-69 %) and 2 µm band depths (4-34 %), except for spectral group 4 and Sierra Gorda 009 (spectral group 5) with 2 µm band depths < 4%. Those spectral groups are compared to asteroid spectra showing similar silicates absorption bands: Q-, O-, R-, K-, V-types and S-complex.

For the first spectral group, meteorite powdered samples have HRM to C-complex asteroids (Table 2). All three have HRM with Cgh- and Ch-types. NWA 11750 and NWA 12957 also have HRM with B-type asteroids from the C-complex. Chwichiya 002 and EM 200 have HRM with Xc- and Xe-types from the X-complex. The difficulty in distinguishing C- and X-complexes is explained in §2.3.3.

The second spectral group has an HRM with C-complex as well. Aydar 003 have an HRM with Cg-, Cgh and Ch-types. NWA 11086 have an HRM to C-, Cg-, and Xk-types. NWA 5958 has HRM with Ch-, Cgh- and B-types. Finally, EM 100 have an HRM with C-type (Table 2). For the third spectral group meteorite powders have a more variable HRM. The QM was with the X-complex (§4.1), but the HRM is with X-complex as well as with C-complex (Table 2). LoV 200, Sierra Gorda 009 and Dho 2066 have an HRM with X-complex and C-complex asteroids. NWA 8781, Acfer 094 and Kakangari powdered samples have an HRM with C-complex. NWA 12474 has an HRM with B-type asteroids. For the fourth spectral group, the powdered samples QM was best with K-type asteroids. Both meteorites have an HRM to K-type, but also S-complex asteroids (Table 2). Indeed, in band position and depth, the S-complex asteroids are similar to K-type asteroids. The shape of the bands is not a parameter considered in the quantitative parameter comparison, but the meteorites of this spectral group have sharp V-shaped absorption bands, characteristic of K-type asteroids, showing that the QM is necessary for some spectra. In the fifth spectral group, powdered samples match S-complex asteroids, more precisely S- and Sr-types asteroids, as observed with the QM (Table 2). Although, NWA 12334 has an HRM with Sq-type, as well as Q- and K-types. The sixth spectral group also has HRM with S-complex asteroids, as observed with QM (Table 2). JaH 846 has an HRM with Sr-type. NWA 12480 has an HRM with V-type and R-type asteroids. The QM was with Sv-type asteroids (Fig. 6) the intermediate spectra between V- and S-types asteroids.

Finally, we calculated the Closest Matching Asteroids (CMA), which represent the closest asteroid to a meteorite sample in all parameter plots. This is calculated by minimizing the norm of the vector defined by a meteorite (M) to an asteroid (A) : $\|\overrightarrow{MA}\| = \sqrt{(\mathrm{Px}M - \mathrm{Px}A)^2 + (\mathrm{Py}M - \mathrm{Py}A)^2}$, with Px the spectra parameter chosen on the x-axis and Py the spectral parameter on the y-axis (Table 2).

The CMA for the first spectral group are B-, Cg- and Ch-types asteroids (Table 2). This is in line with the QM and HRM that linked those meteorites to the C-complex. The second spectral group's CMA are B-, C-, Cg- and Ch-types, also in line with the C-complex linked to those meteorites (Table 2). The third spectral group of meteorites have a CMA of C-complex as well as X-complex, just as seen with the HRM (Table 2). LoV 200, Dho 2066 and NWA 8781 have a CMA of X-type (Xc-, Xe- and Xk-types). Acfer 094, Kakangari, Sierra Gorda 009 and NWA 12474 have a CMA of C-complex asteroids. The fourth meteorite spectral group all have QM to K-type asteroids (Fig. 6), but the CMA are K- and Sq-types asteroids (Table 2). For the fifth spectral group of meteorites, the CMA are S-complex asteroids, especially S- and Sv-types (Table 2). Finally, for the sixth spectral group, the CMA are R-, V- or S-types (Table 2). NWA 12334's CMA are Sq-type asteroids. NWA 12480's CMA are V- and R-types asteroids, similar to the HRM. Finally, JaH 846's HRM are Sr-types asteroids, and its CMA are R-type asteroids, which are one end-member or the Sr-type asteroids, an intermediate between S- and R-types.

## 5.3 Effect of texture on matching

In addition to the comparison of meteorite powdered samples, we also compared raw and polished section meteorite spectra to asteroid spectra. While we discussed how sample preparation affects the reflectance spectra in §4.2, we know discuss how it impacts the connection to potential parent bodies.

### 5.3.1 Raw meteorites samples

As a general behavior, raw meteorite samples were observed to have bluer spectra than meteorite powders (explained §4.2). Consequently, several of our asteroid matches are impacted by sample preparation. In particular, many associations to spectral types were found to be B-type for raw meteorites, while being C-type or even S- and X-types for meteorite powders (Table 2). In the case of meteorite powders with association to S-complex, the spectral match generally remained within the S-complex in the case of raw samples (Table 2). The two studied R chondrites were qualitatively associated to K-types asteroids for powder spectra, and raw spectra as well. For the quantitative comparison of meteorites to asteroids, some changes were found between raw samples and powders, with spectra from raw samples tending to be associated to S-complex or Q-type asteroids.

### 5.3.2 Polished sections

Polished sections are widely used in Earth Sciences. However, this non-natural sample preparation may induce artifacts if one wants to compare with remote observations of small bodies. For the QM, polished samples were matched similarly to powder and raw samples (Table 2). One exception is NWA 11750, which was matched to Cg-type for powdered sample, while matched to Xc-type when prepared as a polished section. In the case of the quantitative comparison, polished section spectra usually match the same asteroid types as either raw or powdered sample preparation. Exceptions can be found, such as Kakangari and Dho 2066, where all three types of sample preparations show different matches for the quantitative comparison. We conclude that the spectral study of polished section can be useful, but should best be used in complement to other sample preparations, or when other types of samples are unavailable.

## 5.4 Potential parent bodies

### 5.4.1 Bus DeMeo taxonomy matching

After the QM, HRM and CMA of meteorites to asteroids was done in this study, we also submitted our meteorites to the online Bus DeMeo taxonomy matching tool (http://smass.mit.edu/cgi-bin/busdemeoclass-cgi), to see if it provided similar results. The Bus DeMeo taxonomy matching is based on powdered samples, but we also tried matching data from raw samples and polished sections. Even if most of the results are comparable to our results, Bus DeMeo taxonomy matching showed limitations. Firstly, almost all spectra from the raw samples are falsely matched to B-type asteroids, due to their bluer slopes (see §4.2). Most of the raw sample spectra matched to B-type asteroids show silicates absorption bands, in contrary to B-type asteroids spectra, invalidating those matches. Also, some of the meteorites showing silicate absorption bands are wrongly matched to C- or X-complexes showing slight to no silicate absorption bands, and for some samples no match could be found.

### 5.4.2 Potential parent bodies of this study

In our study, the first spectral group is composed of three ungrouped type 3.00 carbonaceous chondrites and one C3 carbonaceous chondrite that have a high proportion of matrix (Table 1). With the sample preparation influencing the spectra, we matched them to either Cg-type or B-type asteroids (Table 2). Interestingly, both of those asteroid types have not been linked to any group of meteorites yet (Vernazza and Beck, 2017). This match could demonstrate that those three meteorites create a new spectral group of meteorites with the association of a potential parent body. In the second spectral group, Aydar 003 and NWA 5958 can be matched with Ch-type asteroids (Table 2), in line with CM meteorites that have been linked to Ch- and Cgh-types asteroids (Vernazza and Beck, 2017). EM 100 and NWA 11086 also have some similarities with CM-like spectra, but cannot be matched to Ch-type asteroids. Indeed, they show higher albedo and spectral

VIS-NIR slopes (Fig. 7). Thus, they may originate from an unidentified yet slightly hydrated carbonaceous asteroid type.

The third spectral group of meteorites are linked to the X-complex (Fig. 6c). Indeed, LoV 200, Dho 2066 and NWA 8781 are linked to Xe-type or Xk-type asteroids. Xe-type asteroid have not been linked to any meteorite groups yet (Vernazza et al., 2016) and NWA 8781, Dho 2066 and LoV 200 could be potential samples of this asteroid type. Kakangari and Sierra Gorda 009 are both linked to X-complex as well as C-complex asteroids. They could therefore be matched to both C- and X-complexes (Table 2). For Acfer 094 and NWA 12474, we are not able to link those meteorites to any types of asteroid. Using the three sample preparations, we can see that their matching is not straightforward (Table 2) and would require further investigations, although they would probably originate from asteroids resembling X-complex. NWA 12474 is indeed matched to B-type due to its bluer slope but shows slight silicate absorption bands, invisible on B-type asteroids. In this subgroup, chondrites of various petrology match the same asteroid types, such as Kakangari, Sierra Gorda009, Dhofar 2066 and NWA 8781. Those chondrites probably do not originate from a common parent body, as their petrography are too different. Nonetheless, they do probably come from parent bodies displaying similar VIS-NIR spectra as the X-complex. This demonstrate a limit of the spectral matching, as some mineralogy, although different, result in similar spectra thus a similar match.

The fourth spectral group (Rumuruti chondrites) is matched to K-type asteroids (Fig. 6d). Rumuruti meteorites were previously linked to the A-type asteroids (Sunshine et al., 2007), but K-type shows a better match in our analysis, as the A-type reflectance spectrum shows too deep absorption bands and too high VIS-NIR slope compared to our meteorite's spectra.

The fifth spectral group is composed of LoV 051, Coolidge and SaH 00177. They are linked to S-, Sr- or Sv-type asteroids (Table 2). This suggests that the S-complex of asteroids is not only composed of ordinary chondrites or primitive achondrites, but may also host metamorphosed carbonaceous chondrites. SaH 00177, studied only as polished sections, also match S-, Sr, and Sv-types asteroids. The sixth spectral group is composed of two metamorphosed ordinary chondrites and two primitive achondrites. We link the three samples to asteroid spectra with deep absorption bands, such as Sv-, Sq- and Sa-types asteroids. The acapulcoite matches Sv-type asteroids, NWA 12334 (LL6-an) matches S- or Sq-type asteroids, and JaH 846 is linked to Sr- or Sv-type asteroids. Finally, NWA 6592 and Mulga (west), only studied as polished sections, are good matches to Sa- or Sq-types asteroids.

# 6 Conclusion

We studied 25 meteorites, mostly ungrouped chondrites or meteorites from rare groups. We first separated them into six meaningful petrographic groups. The same meteorites were studied by VIS-NIR reflection spectroscopy, to compare their spectra to asteroid reflectance spectra. The spectra obtained of all meteorites were grouped into 6 spectral groups. Those spectral groups mirror the petrographic groups, with only a few exceptions (Acfer 094, NWA 8781, EM 200, NWA 12334, and JaH 846). VIS-NIR spectra are therefore a good proxy for the overall petrography, and as such a meaningful tool to match meteorites and potential parent asteroids. In order to better match real asteroid surfaces that can be dusty and/or rocky, we measured raw samples in additions to powdered samples. The two types of samples have different spectral characteristics. Raw samples show bluer VIS-NIR slopes than powders, especially for carbonaceous chondrites. We also compare the powders' spectra to polished sections spectra, as polished sections are widely available for study. We show that polished section spectra can be good equivalent to raw sample spectra, with similar spectral parameters and matching to asteroids. The matching of meteorite samples to asteroids were made by qualitative as well as quantitative matching of different parameters (bands depths, bands positions and slopes at 1 and 2 µm). Interestingly, some ungrouped meteorites are matched to asteroid types which were not linked to any grouped meteorites before this study. Indeed, the three ungrouped carbonaceous C3.00 chondrites (spectral group 1), representing very pristine carbonaceous chondrite material, are matched to Cg-type or B-type asteroids. In addition, three slightly metamorphosed and aqueously altered ungrouped carbonaceous chondrites from the spectral group 3 are good matches to Xe-type asteroids, which were also not matched to any other grouped meteorites before. Some meteorites from the spectral group 2, showing CM-like petrography, resemble Ch-type asteroids, but showing more hydration bands than the asteroid spectra. This shows that those samples must originate from an unidentified yet slightly hydrated carbonaceous asteroid type. Those samples should be further investigated, as the current mission OSIRIS-REx showed that the asteroid (101955) Bennu should be a CM-like asteroid with more hydration than usual CM chondrite spectra (Lauretta et al., 2019). In addition, some metamorphosed carbonaceous chondrites, from our spectral group 5 and 6, matched with the S-complex asteroids, usually presented as parent bodies of ordinary chondrites and primitive achondrites, suggesting that the S-complex of asteroids may also host metamorphosed carbonaceous chondrites. We also compared our matching results to the Bus DeMeo taxonomy online program. Their matches were similar to ours for the powdered samples, unlike for the raw samples where the Bus DeMeo program wrongly matched the samples mainly to B-type asteroids. This emphasizes our conclusion that the use of only powdered samples in parent body research may be misleading, as asteroid's surfaces not only display regolith, but also large blocks (e.g., asteroid (101955) Bennu). In conclusion, the study of ungrouped chondrites and other rare meteorites reveal new links to yet unmatched asteroids, exemplifying the importance of the study of this material which represent a large fraction of the diversity of the meteoritic material available for study, and could sample many unmatched asteroids. As such, these meteorites give insight into the composition and mineralogy of unstudied asteroids.

# 7 Acknowledgment

This work was funded by the European Research Council under the H2020 framework program/ERC grant agreement no. 771691 (Solarys). The help of Lydie Bonal is acknowledged for importing the spectra in the SSHADE database

**Figures Legends** :

Figure 1 : BSE images of chondrites from petrographic group A. a: Chwichya002 (C3.00-ung); b : NWA 11750 (C3.00-ung); c : NWA 12957 (C3.00-ung); d: Acfer 094 (C2-ung). Scale bar is identical for all images.

Figure 2: BSE images of chondrites from petrographic group B: a : Aydar 003 (CM1/2); b : EM 100 (C2-ung); c : NWA 11086 (CM-an); d : EM 200 (C3); e : NWA 5958 (C2-ung); f: NWA 8781 (C-ung). Scale bar is identical for all images.

Figure 3: BSE images of chondrites from petrographic group C: a: LoV 200 (CH3); b : Dho2066 (C-ung); c: Kakangari (K3); d: NWA 12474 (CR3); e: SG 009 (chondrite-ung). Scale bar is identical for all images.

Figure 4: BSE images of chondrites from petrographic group C: a : Awsserd (R4) ; b: NWA 12472 (R3). BSE images of chondrites from petrographic group E: c: LoV 051 (C3-ung) ; d : Coolidge (C4-ung) ; e : Sah00177 (C3/4-ung) ; f: JaH 846 (OC3). Scale bar is identical for all images.

Figure 5: BSE images of meteorites from petrographic group F: a : Mulga West (C5/6-ung) ; b: NWA 12334 (LL6-an) ; c: NWA 12480 (Acapulcoite) ; d: NWA 6592 (Lodranite). Scale bar is identical for all images.

Figure 6 : Meteorite spectra (solid line) compared to asteroid spectra (dashed lines). Spectra were normalized at 0.55 and then offset for clarity. Number 1-6 are spectral group of meteorites. “L” is for leached powdered samples; ACA: acapulcoite; Chond-ung : chondrite ungrouped.

Figure 7 : Equivalent geometric albedo of meteorites and asteroid complexes. Number 1-6 are spectral groups of meteorites. Equivalent albedo from reflectance at 550nm, phase angle g=30°, calculated with factor from Figure 3 of Beck et al., (2021). Geometry albedo of asteroids families from DeMeo and Carry (2013). ACA: Acapulcoite; Chond-ung : ungrouped chondrite.

Figure 8 : Reflectance for different sample preparations. A: Sierra Gorda 009 samples, from petrographic group C and spectral group 3; B: Chwichiya 002 sample, from petrographic group A and spectral group 1. P.S. : polished section.

Figure 9 : Slopes ($nm^{-1}$) at 1 (cold colors) and 2 µm (warm colors) for different sample preparation for all spectral groups.

Figure 10: Visible-near-infrared reflectance spectra of studied samples (solid line) compared to spectra of selected groups of meteorites (dashed lines). Spectra were normalized at 0.55 µm and then offset for clarity. Number 1-6 are spectral group of meteorites. No comparison for the spectral group 4 was made, as meteorites in this spectral group belong to the well-established R chondrite group. “L” is for leached powdered samples; ACA : acapulcoite; Chond-ung : chondrite ungrouped.

Figure 11: Plots of spectral parameters for NWA 11750 (C3.00-ung), belonging to petrographic group A and spectral group 1. Asteroid Highest Recurrent Match (HRM) is Cg-type. Yellow ellipses correspond to bulk sample spectrum data; Green ellipses correspond to powder sample spectrum data; Red ellipses correspond to polished section spectrum data. Chond-ung : chondrite ungrouped; P.S.: polished section.

Figure 1: Plots of spectral parameters of LoV 051 (C3-ung), belonging to petrographic group E and spectral group 5. Asteroid Highest Recurrent Match (HRM) is Sr-type. Yellow ellipses correspond to bulk sample spectrum data; Blue ellipses correspond to leached powder sample spectrum data; Red ellipses correspond to polished section spectrum data. Chond-ung : chondrite ungrouped; achondrite-prim : primitive achondrites; P.S.: polished section; Powder-L : powdered leached sample.

Table 1 : Petrographic characteristic of meteorites.

Table 2 : Asteroids matching to meteorites.

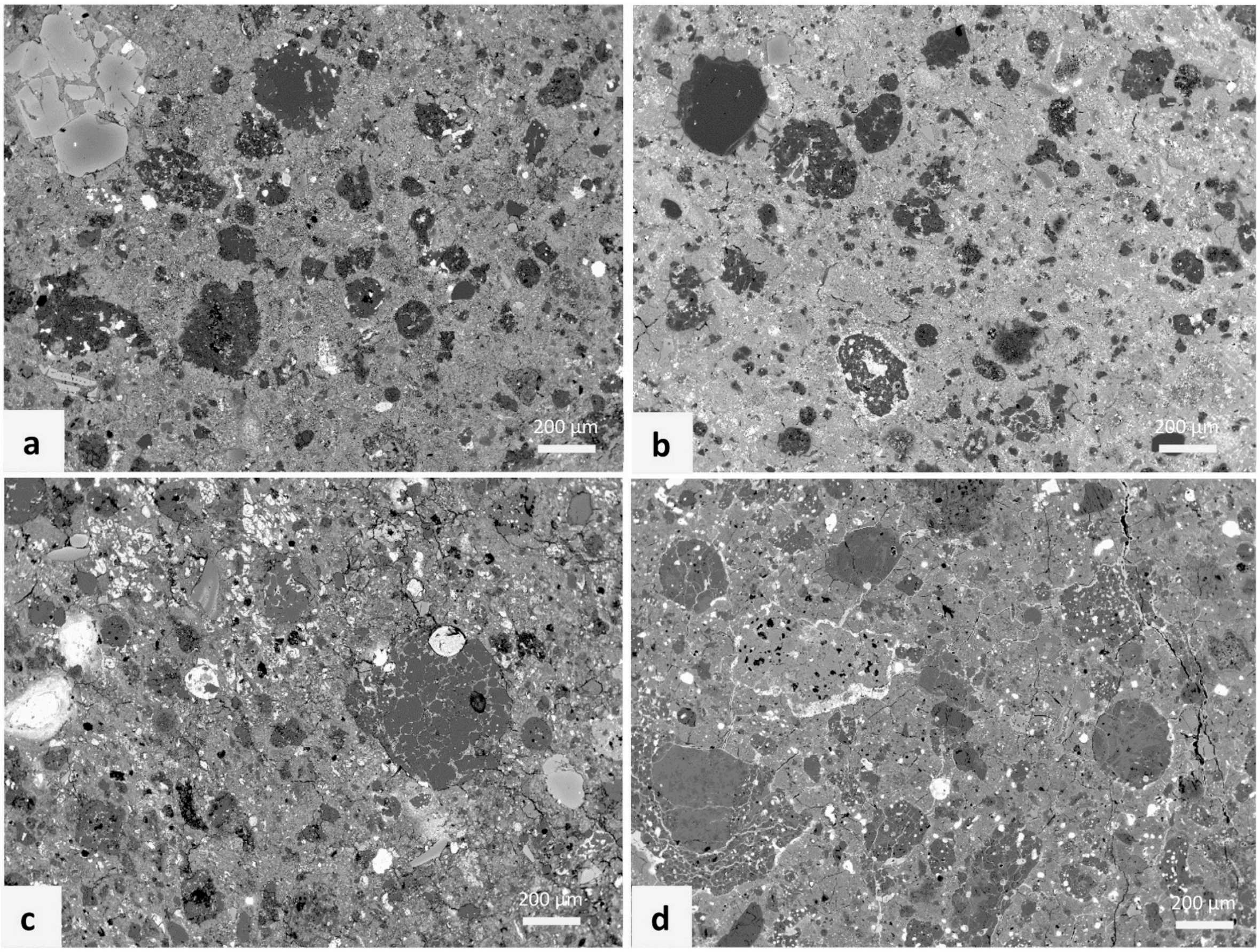

FIGURE 1

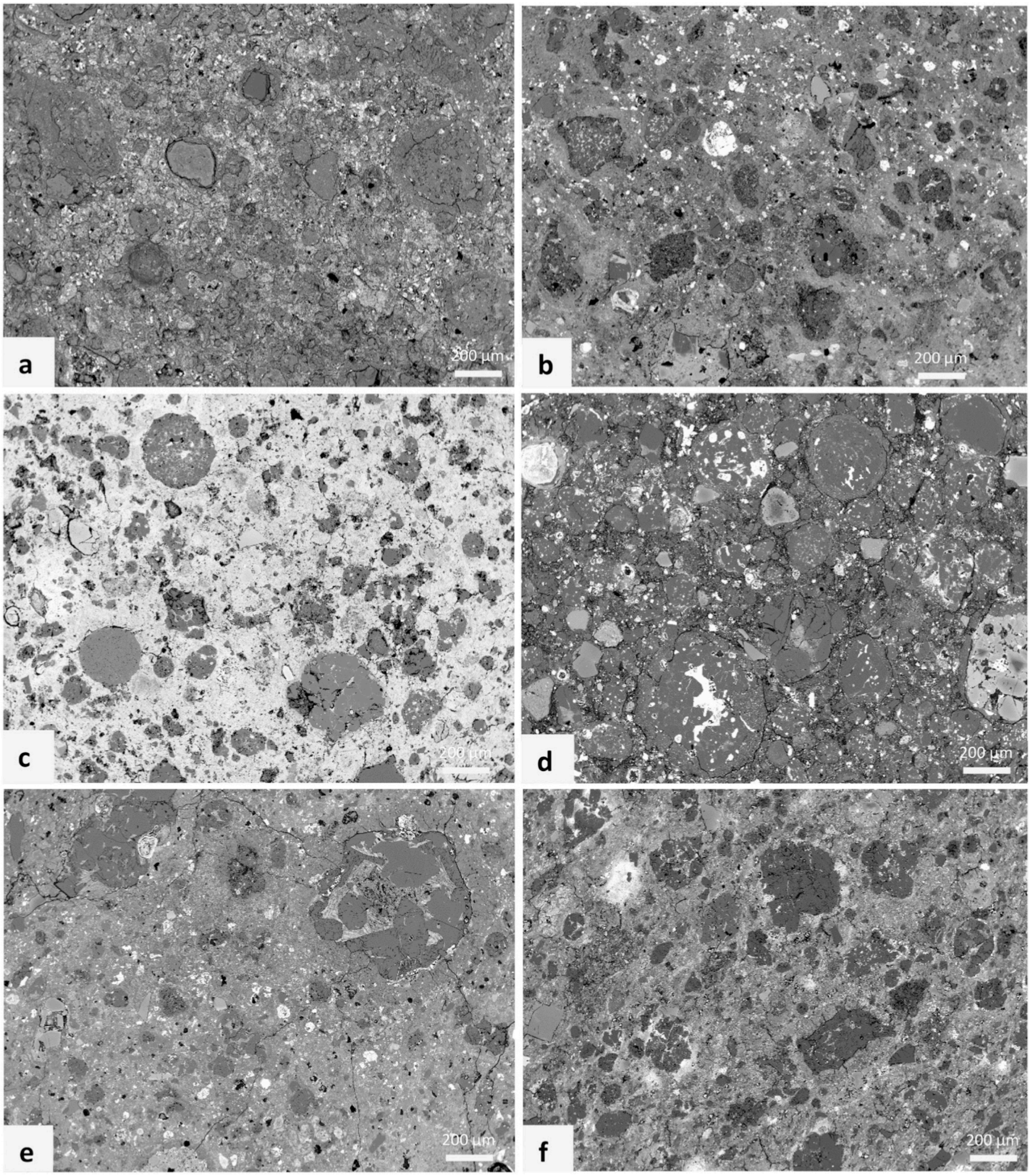

FIGURE 2

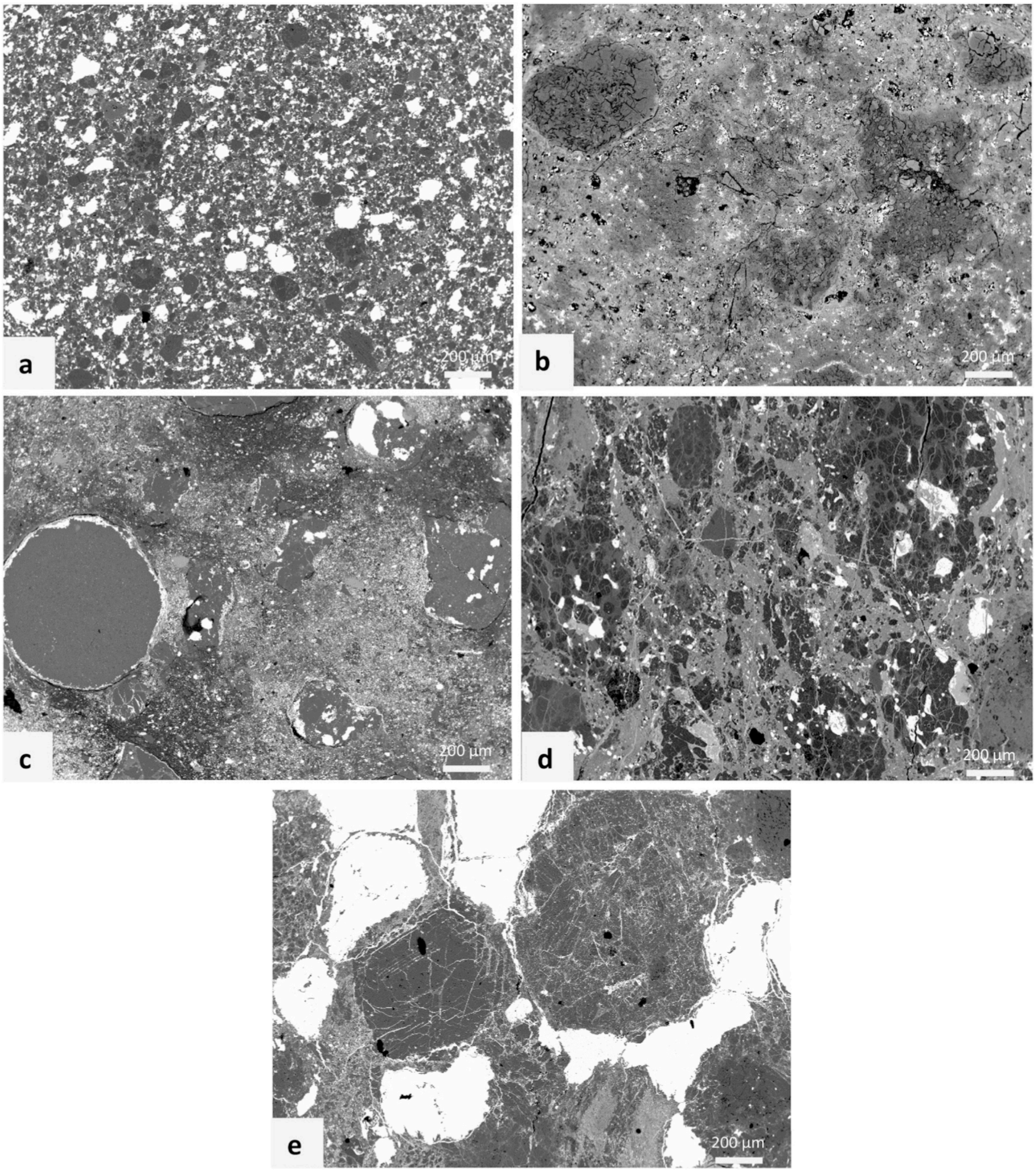

FIGURE 3

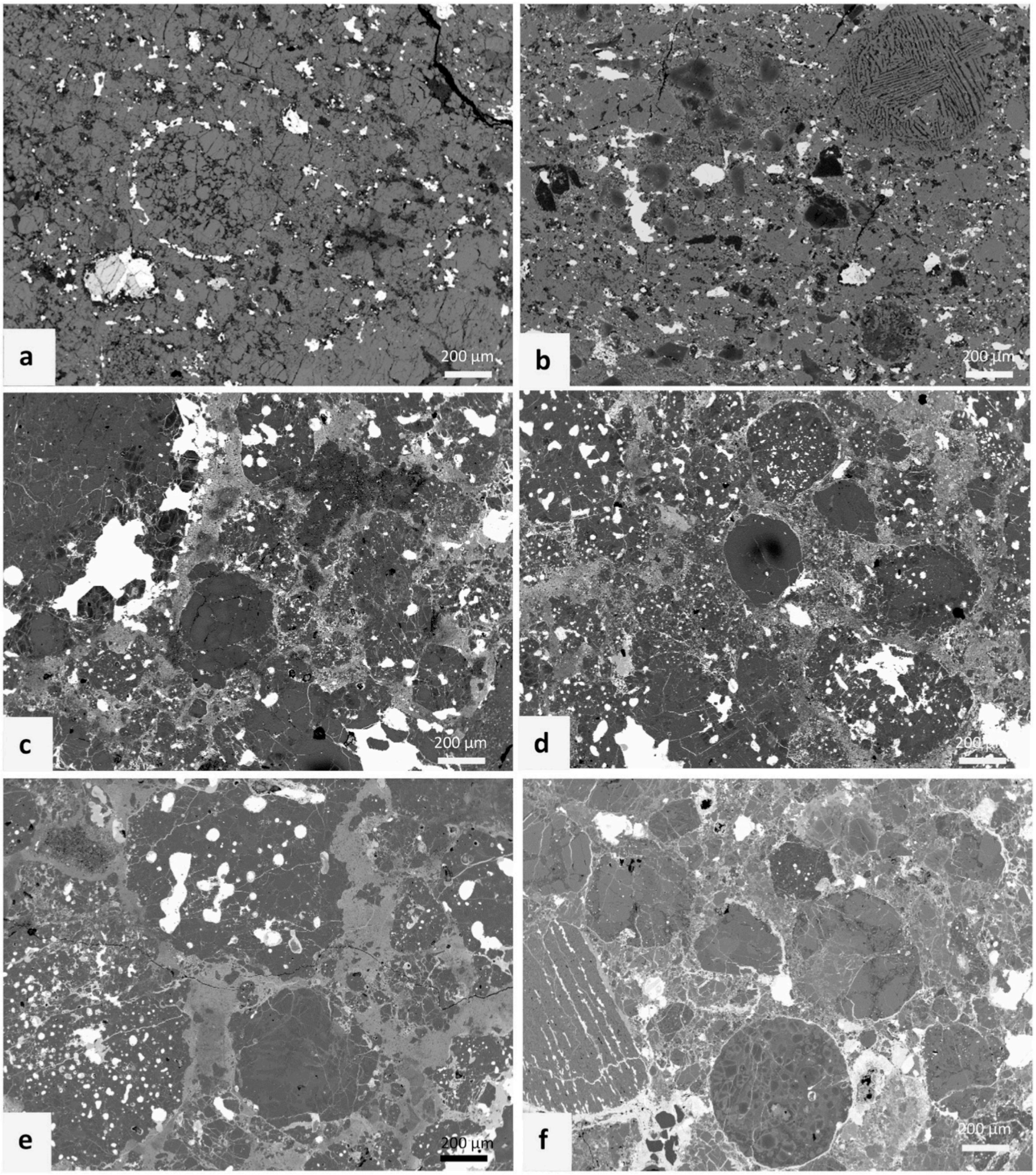

FIGURE 4

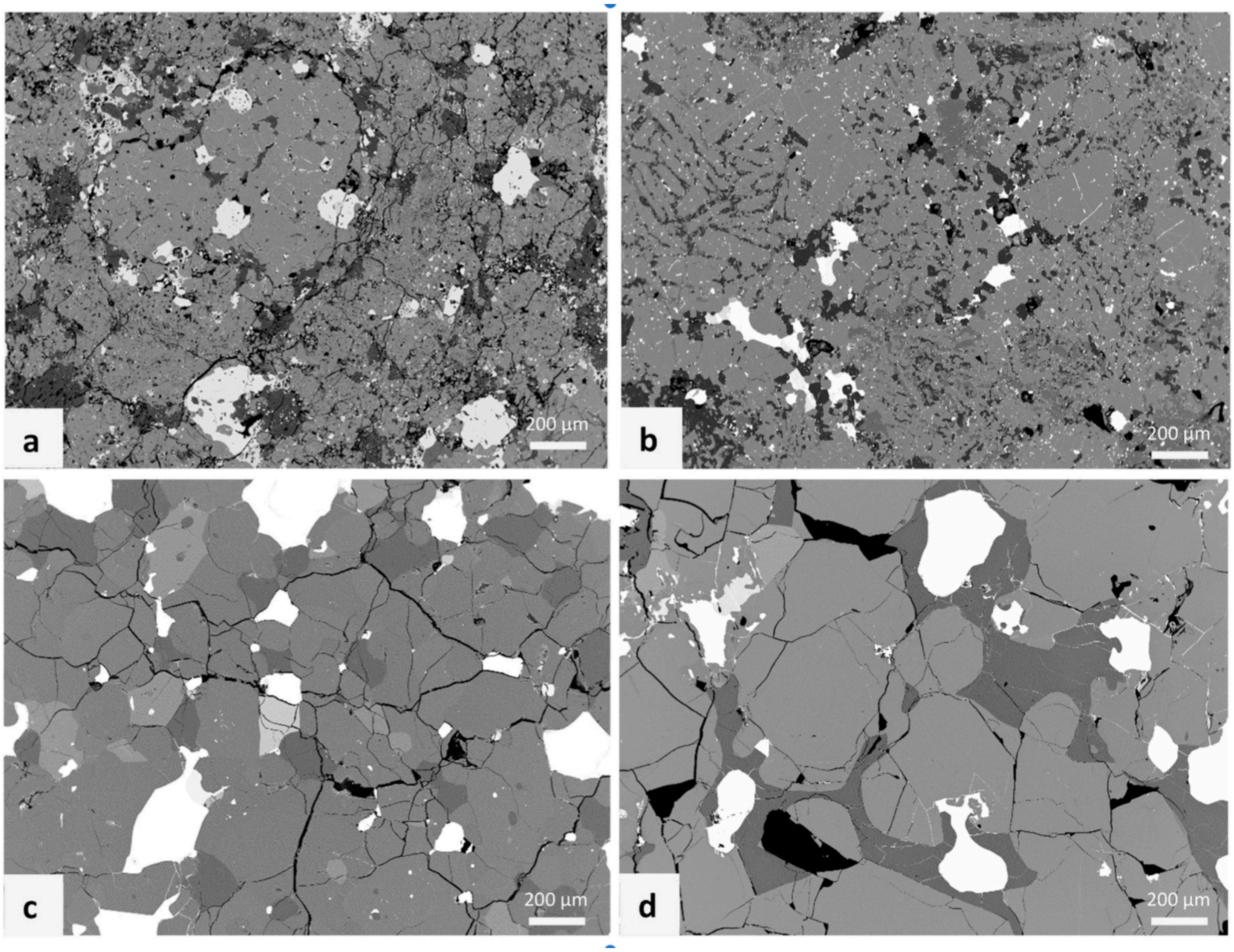

FIGURE 5

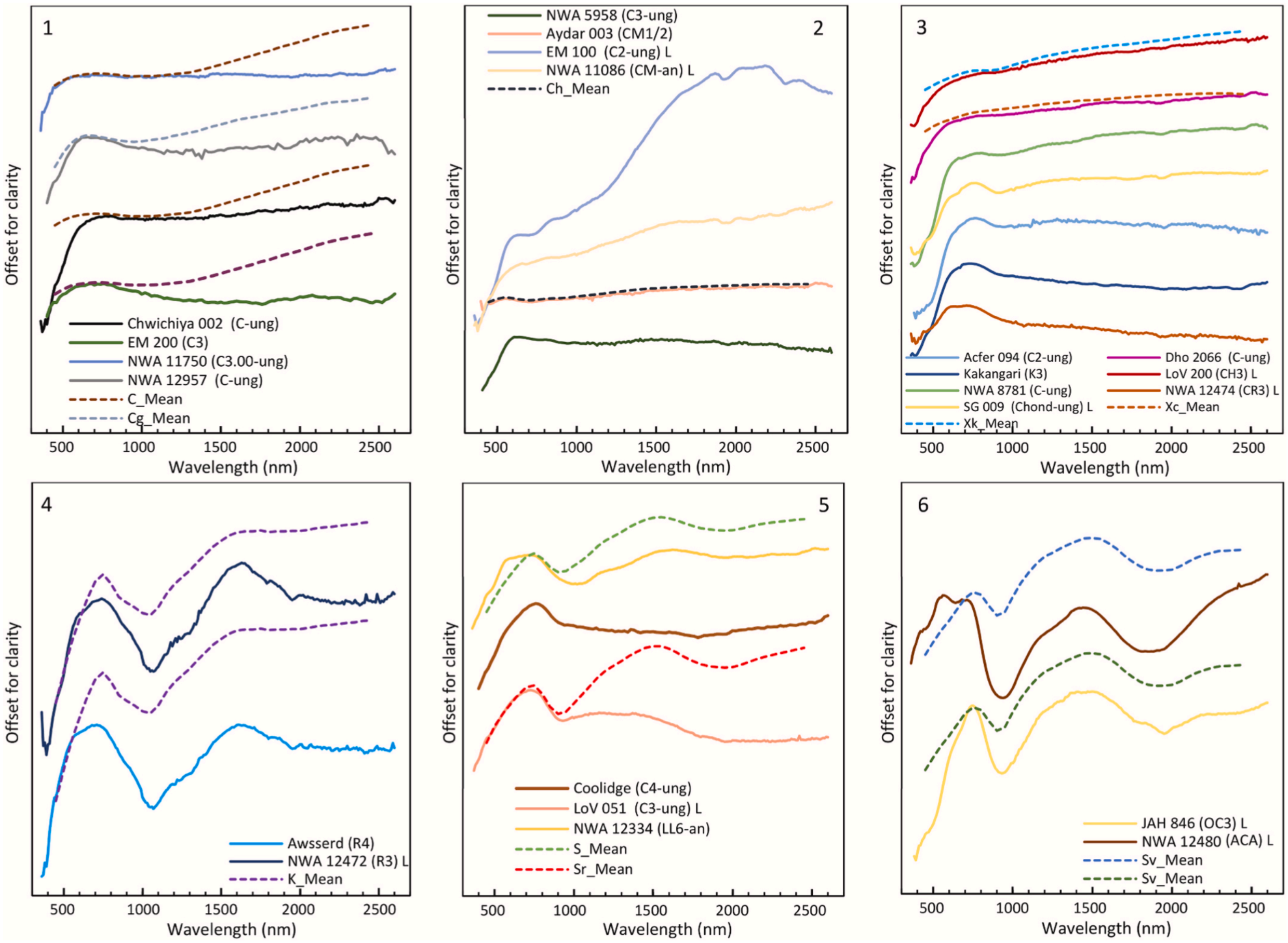

FIGURE 6

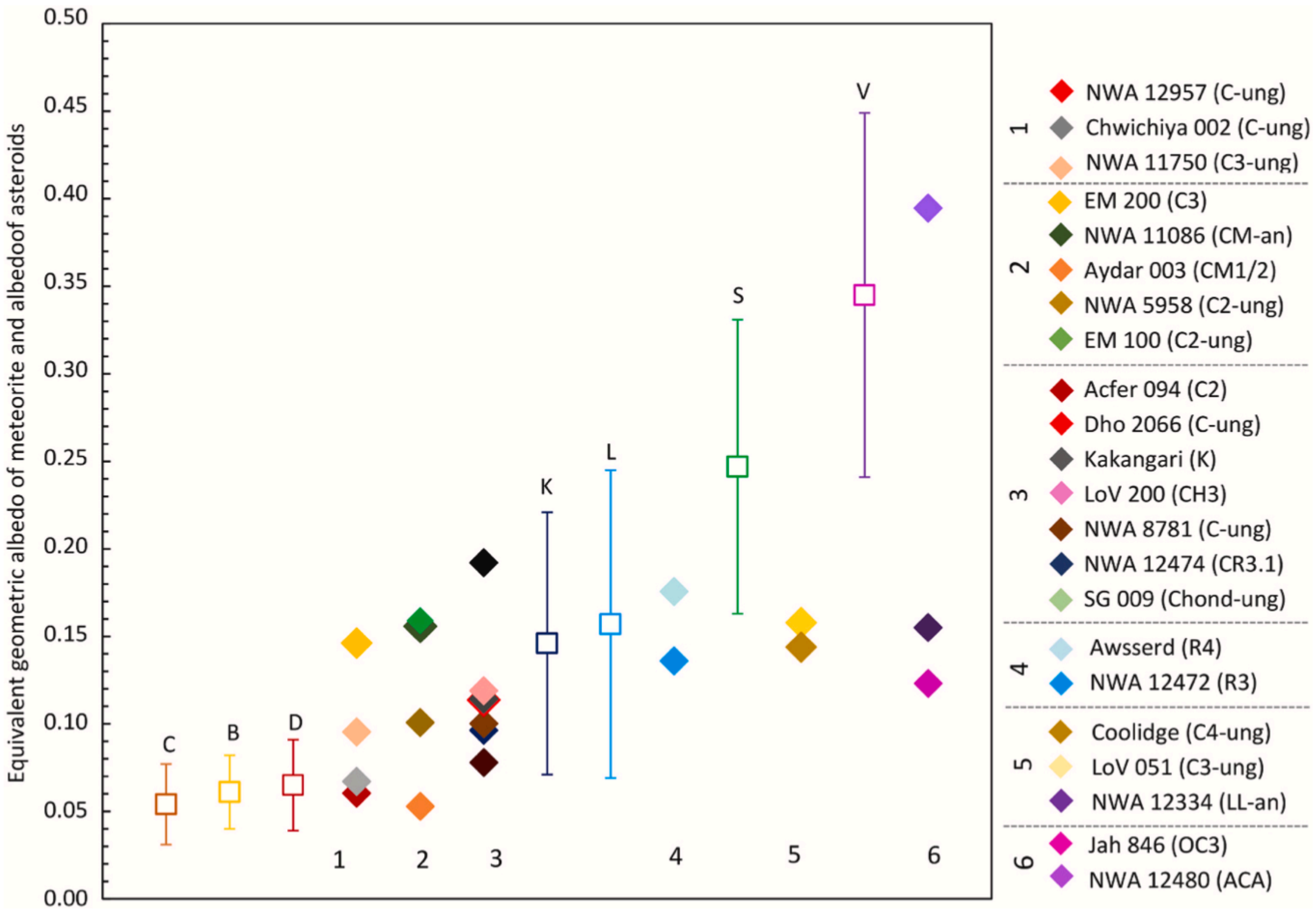

Equivalent geometric albedo of meteorite and albedoof asteroids
0.00
0.05
0.10
0.15
0.20
0.25
0.30
0.35
0.40
0.45
0.50
C
B
D
K
L
S
V
1
2
3
4
5
6
1
NWA 12957 (C-ung)
Chwichiya 002 (C-ung)
NWA 11750 (C3-ung)
2
EM 200 (C3)
NWA 11086 (CM-an)
Aydar 003 (CM1/2)
NWA 5958 (C2-ung)
EM 100 (C2-ung)
3
Acfer 094 (C2)
Dho 2066 (C-ung)
Kakangari (K)
LoV 200 (CH3)
NWA 8781 (C-ung)
NWA 12474 (CR3.1)
SG 009 (Chond-ung)
4
Awsserd (R4)
NWA 12472 (R3)
5
Coolidge (C4-ung)
LoV 051 (C3-ung)
NWA 12334 (LL-an)
6
Jah 846 (OC3)
NWA 12480 (ACA)

FIGURE 7

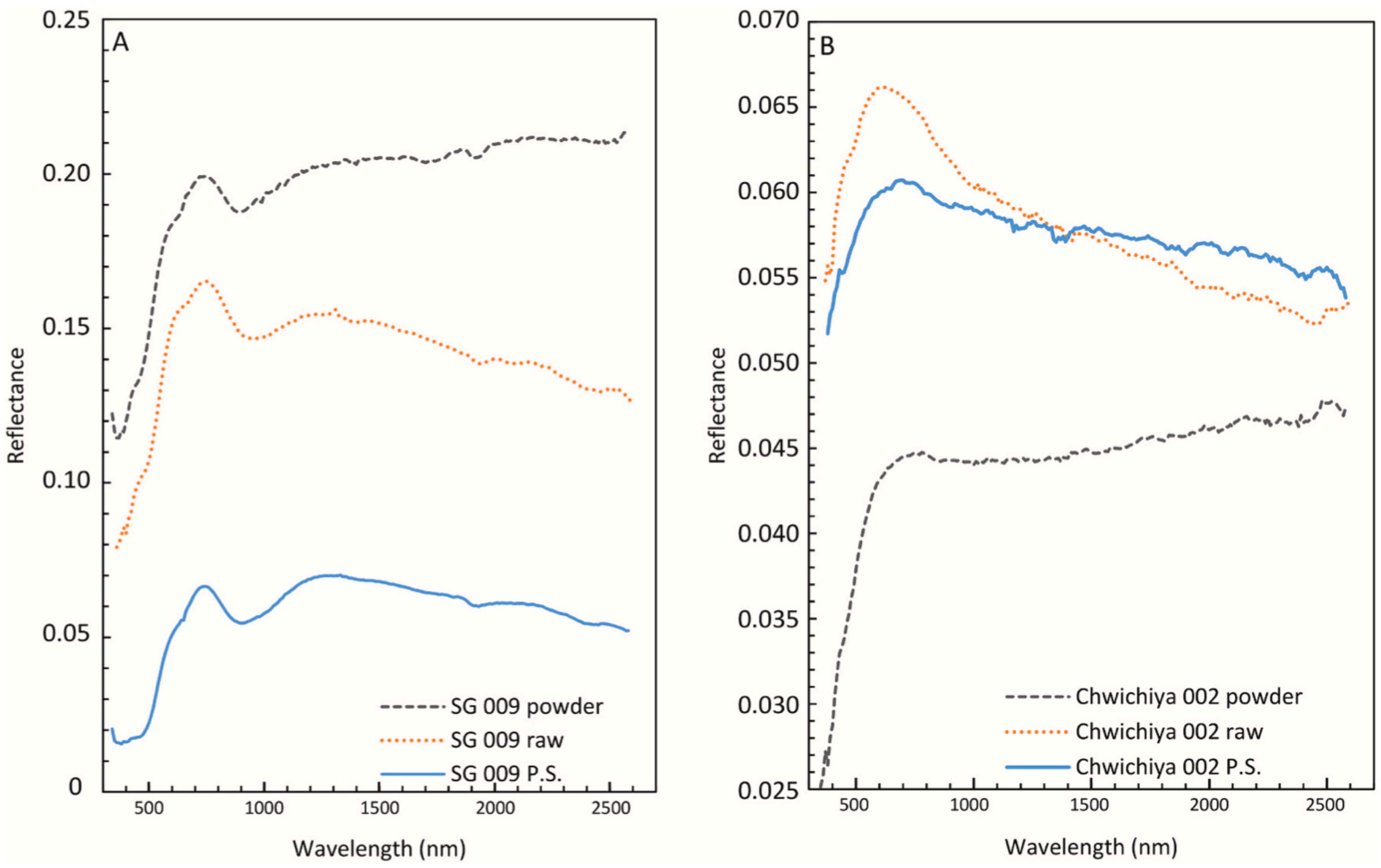

FIGURE 8

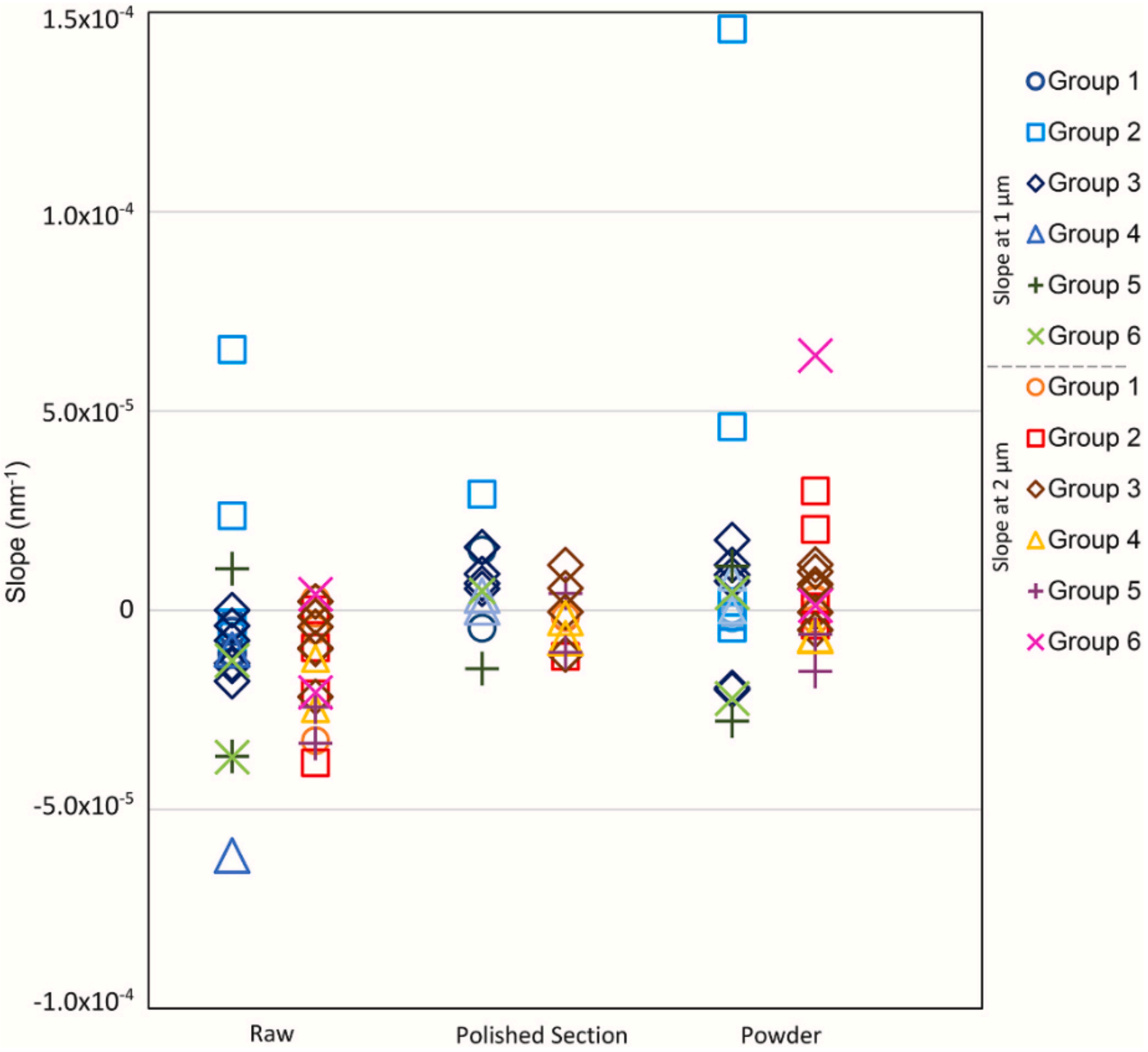

FIGURE 9

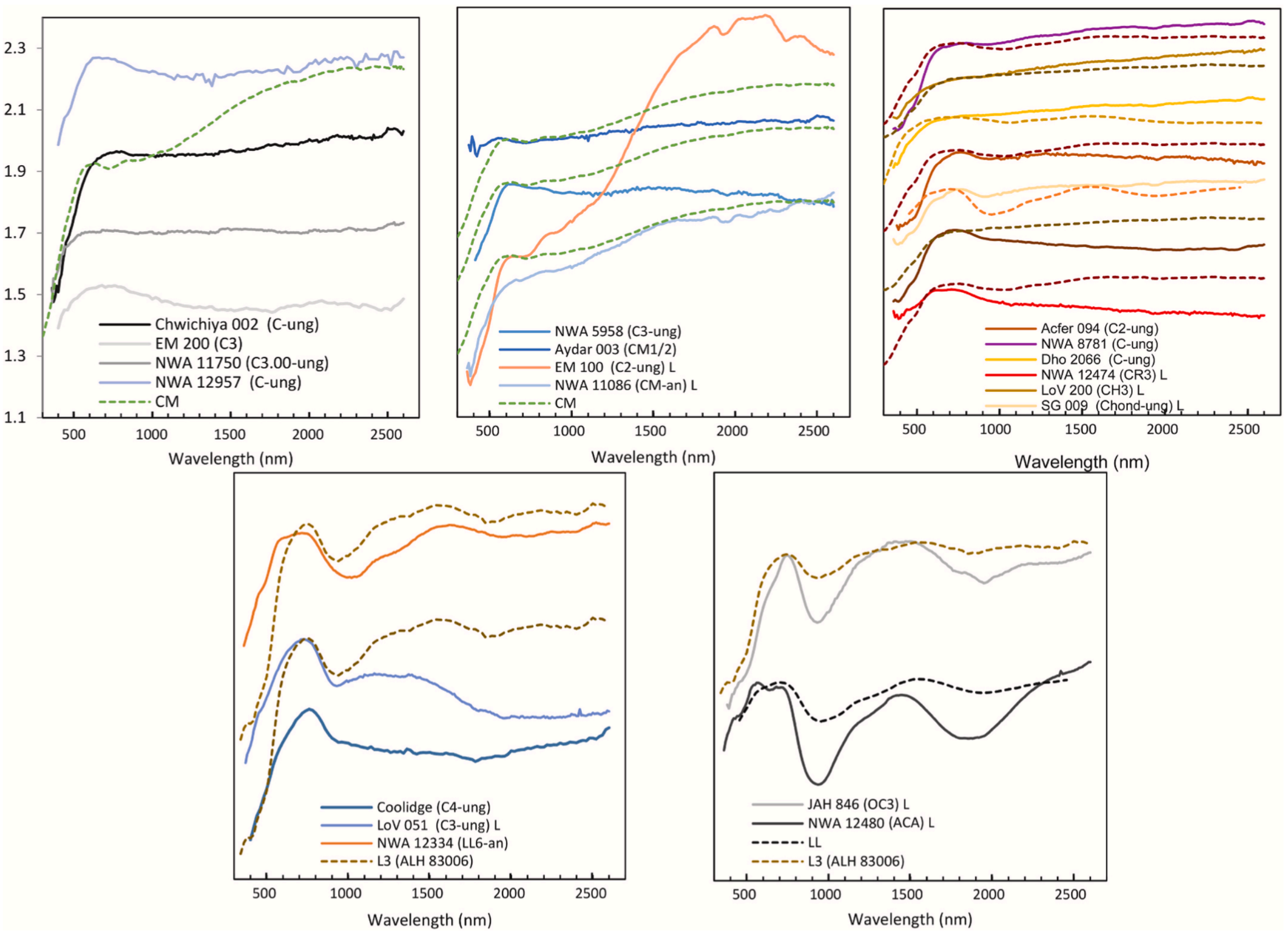

Chwichiya 002 (C-ung)
EM 200 (C3)
NWA 11750 (C3.00-ung)
NWA 12957 (C-ung)
CM
Wavelength (nm)
NWA 5958 (C3-ung)
Aydar 003 (CM1/2)
EM 100 (C2-ung) L
NWA 11086 (CM-an) L
CM
Wavelength (nm)
Acfer 094 (C2-ung)
NWA 8781 (C-ung)
Dho 2066 (C-ung)
NWA 12474 (CR3) L
LoV 200 (CH3) L
SG 009 (Chond-ung) L
Wavelength (nm)
Coolidge (C4-ung)
LoV 051 (C3-ung) L
NWA 12334 (LL6-an)
L3 (ALH 83006)
Wavelength (nm)
JAH 846 (OC3) L
NWA 12480 (ACA) L
LL
L3 (ALH 83006)
Wavelength (nm)

FIGURE 10

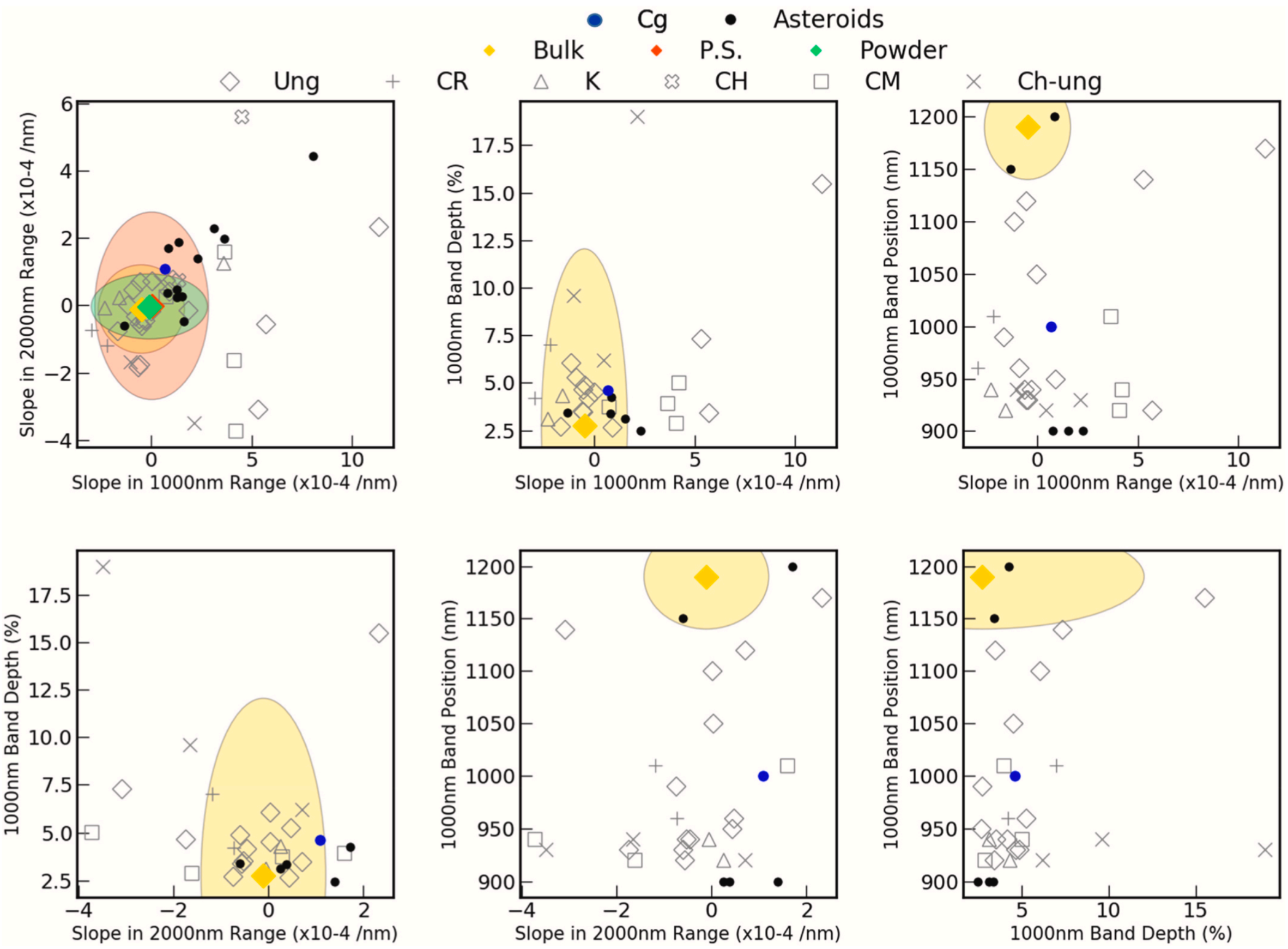

Cg
Asteroids
Bulk
P.S.
Powder
Ung
CR
K
CH
CM
Ch-ung
Slope in 2000nm Range (x10-4 /nm)
Slope in 1000nm Range (x10-4 /nm)
1000nm Band Depth (%)
1000nm Band Position (nm)
Slope in 2000nm Range (x10-4 /nm)
1000nm Band Depth (%)

FIGURE 11

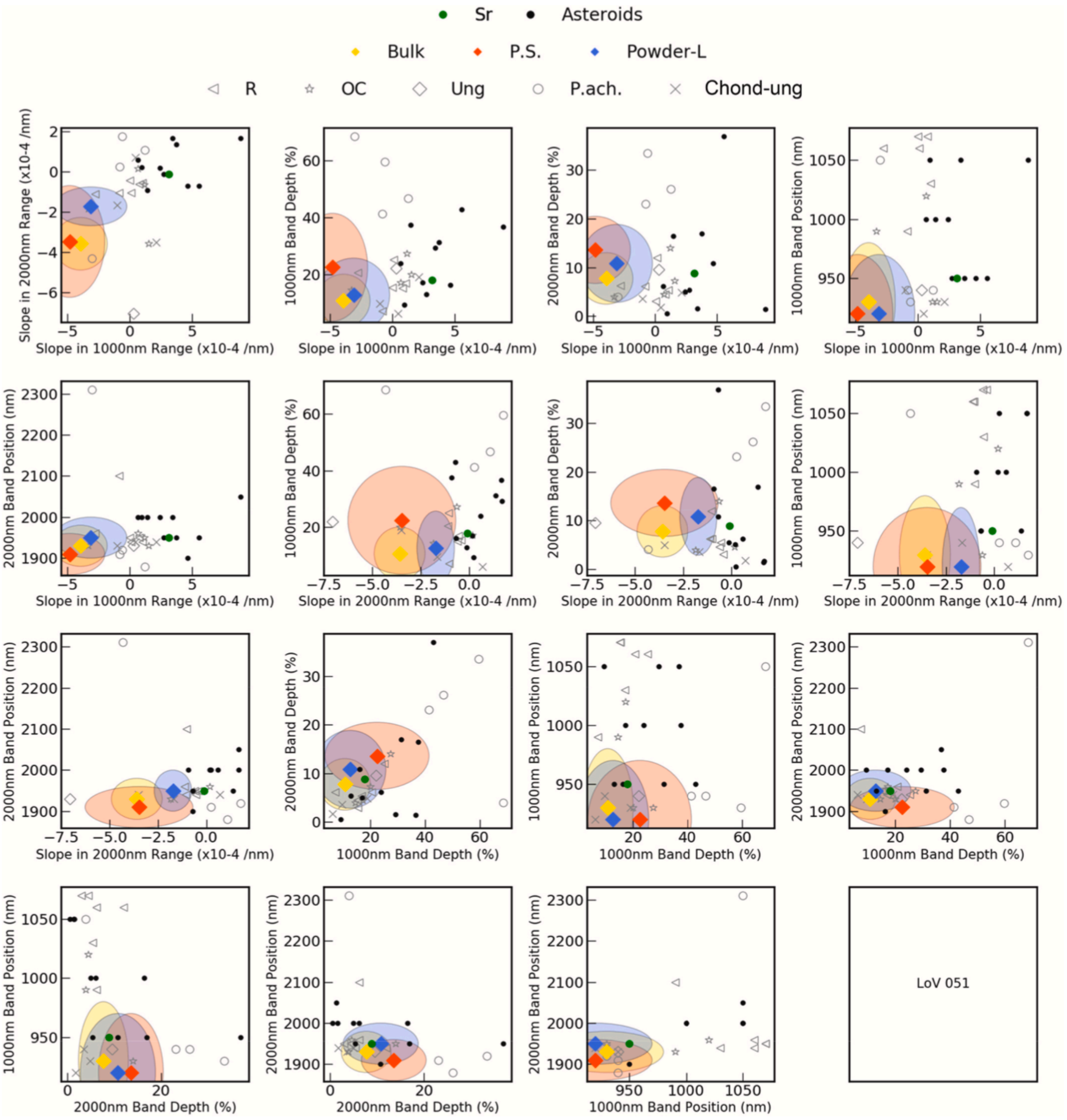
Sr
Asteroids
Bulk
P.S.
Powder-L
R
OC
Ung
P.ach.
Chond-ung
Slope in 1000nm Range (x10-4 /nm)
Slope in 2000nm Range (x10-4 /nm)
1000nm Band Depth (%)
2000nm Band Depth (%)
1000nm Band Position (nm)
2000nm Band Position (nm)
LoV 051

FIGURE 12

**Table 1**
Petrographic characteristic of meteorites.

| Petro Group | Meteorite | Group | Type | Matrix | | Chondrules | | | Weathering | Olivine | | Low-Ca Pyroxene | | | Plagioclase | | Sulfides | Metal | Magnetite | Mag. Susc. | Ms |
|---|---|---|---|---|---|---|---|---|---|---|---|---|---|---|---|---|---|---|---|---|---|
| | | | | (vol %) | n | (vol %) | (μm) | n | | Fayalite (mol %) | n | Ferrosilite (mol %) | n | Wollastonite (mol %) | (mol%) | n | (vol%) | (vol %) | (vol%) | (log χ) (×10$^{-9}$m$^3$/kg) | (Am$^2$/kg) |
| A | Chwichiya 002[a] | C3.00-ung | 3 | 73.4* | 421 | 12.9* | 480 ± 300 | 29 | moderate | 37.6 ± 16.7 | 16 | 3.1 ± 3.0 | 4 | 0.8 ± 3 | – | – | 4.9* | <1.0* | 7.8* | 4.38* | 3.29* |
| A | NWA 11750[b] | C3.00-ung | 3.00 | 74.0 | / | 26.0 | 240 ± 170 | 28 | minimal | 11.9 ± 17.1 | 7 | 3.2 ± 2.2 | 3 | 3.8 ± 1.1 | – | – | 1.9** | | – | 4.03* | 2.36* |
| A | NWA 12957[c] | C3.00-ung | 3 | 63.3* | 667 | 24.8 | 300 ± 200 | 25 | low | 24.3 ± 20.8 | 21 | 8.8 ± 12.7 | 10 | 1.5 ± 0.8 | – | – | 6.6* | 1.0* | 4.1* | 4.44* | 6.19* |
| A | Acfer 094[d] | C2-ung | 2 | 60.9* | 368 | 34.5* | 130 ± 125 | 82 | moderate | 0–55.0 | / | 1–16 | / | 35.5–46.0 | $An_{70-98}$ | / | 1.4* | 0.5* | – | 4.62* | 11.43* |
| B | NWA 5958[c] | C2-ung | 2 | 75.5 | / | 18.9 | 180 ± 100 | 1272 | W1 | 2.2 ± 1.1 | 7 | 2.6 ± 2.6* | 8* | 1.1 ± 0.2* | – | – | 1.0–2.0 | <1.0 | 1.2 | 416* | 2.07* |
| B | Aydar 003[e] | CM1/2 | 1 | 73.5* | 401 | 12.2* | 270 | / | minimal | 1.4–14.9 | 2 | 26.3* | 7* | 1* | – | – | 14.0 4.0* | – | 7.5** | 3.75* | 0.90* |
| B | EM 100[f] | C2-ung | 2 | 71.8* | 339 | 21.2* | 250 | / | moderate | 7.3 ± 12.1 | 23 | 3.7 ± 2.8 | 7 | 0.9 ± 01 | – | – | 5.5* | 1.5* | – | 3.93* | 1.34* |
| B | NWA 11086[g] | CM-an | | 58.0* | 471 | 42.0* | 240 ± 120 | 38 | severe | 14.2 ± 12.2 | 13 | 1.4 ± 0.1 | 2 | 1.1 ± 01 | – | – | 2.1* | 3.0* | – | 3.41* | 0.37* |
| B | EM 200[f] | C3 | 3 | 44.1* | 669 | 35.2* | 130 ± 80 | / | low | 22.7 ± 23.6 | 39 | 1.6 ± 0.5 | 7 | 0.7 ± 0.4 | – | – | 7.6* | 3.6* | 32.6** | 5.08* | 17.10* |
| B | NWA 8781[b] | C-ung | | 55.2* | 458 | 33.0* | 210 ± 60 | 20 | low | 1.1–70.1 | 8 | 1.0–7.0 | 3 | 1.2–0.4 | – | – | 9.2* | 2.6* | – | 4.16* | 2.18* |
| C | Kakangari[h] | K3 | 3 | 66.8 | 238 mm$^2$ | 22.7 | 690 | 119 | absent | 3.2–9.3 | / | 3.3–14.3 | / | – | – | – | 10.5 | 5.8 | – | 4.88* | 15.10 |
| C | NWA 12474[e] | CR3 | 3.1 | 56.0* | 888 | 29.3* | 650 ± 310 | 40 | high | 1.9 ± 0.8 | 5 | 2.4 ± 0.4 | 7 | 1.2 ± 0.4 | $An_{86.4\pm2.7}$ $Or_{0.0\pm0.1}$* | 2* | 4.6* | 2.0* | 8.0* | 4.69* | 11.11* |
| C | LoV 200[g] | CH3 | 3 | 0* | 320 | 66.3* | 50 ± 34* | 24 | minimal | 15.2 ± 13.9 | 8 | 13.1 ± 14.0 | 19 | 2.1 ± 1.8 | – | – | 3.8* | 20.8* | – | 5.40* | 63.50* |
| C | Dho 2066[g] | C-ung | / | 54.3* | 602 | 30.5* | 620 ± 280 | 25 | W1 | 1.7 ± 2.3 | 10 | – | – | – | – | – | 13.2* | <1* | 1.2* | 3.96* | 2.67* |
| C | Sierra Gorda 009[i] | Chondrite-ung | / | 0* | 406 | 74.5* | 1050 ± 577* | 25* | moderate | 0.5 ± 0.1 | 74 | 1.4 ± 0.4 | 42 | 0.9 ± 0.6 | $An_{32.28-95.04}$ $Ab_{4.96-65.99}$ | / | 8.6* | 16.9* | – | 5.28* | 52.40* |
| D | Awsserd[g] | R4 | 4 | 62.6* | 411 | 26.2* | 246 ± 60* | 68* | minimal | 40.9 ± 1.7 | / | 12.2 | / | 44.7 | $An_{7.7}Or_{3.0}$* | 3* | 10.5* | – | – | 3.05* | 0.127* |
| D | NWA 12472[e] | R3 | ≥3.2 | 60.5* | 430 | 27.2* | 326 ± 264* | 37* | high | 37.2 ± 0.1 | 2 | | | | $An_{13.8}Or_{4.2}$* | 6* | 12.3* | – | – | 3.57* | 0.234* |
| E | Sahara 00177[j,o] | C3/4-ung | 3/4 | 22.7* | 449 | 63.3* | 500 | | moderate | 8.0 ± 1.0 | / | 4.1* | 7* | 4.4* | $An_{79.8}Or_{2.1}$* | 7* | 3.6* | 3.6* | – | 5.15* | / |
| E | Coolidge[k] | C4-ung | 3.8–4 | 17.2 | | 67.4 | 610 ± 680 | 52 | very | 14.2 ± 3.0 | 95 | 11.2 ± 11.0 | 49 | 2.7* | $An_{74.8}Or_{3.5}$ | 2 | 2.6 | 3.7 | 0.4 | 4.94* | / |
| E | Mulga (west)[l] | C5/6-ung | 5/6 | / | 191 | 76.4* | / | / | slightly | 32.0 | / | 26 | / | / | $An_{33-64}$ | / | 6.8* | <1.0* | 16.2* | / | / |
| E | LoV 051[m] | C3-ung | 3 | 25.8* | 507 | 55.4* | 940 ± 280 | | minor | 13.7 ± 1.4 | 31 | 9.2 ± 3.3 | 9 | 1.1 ± 0.6 | $An_{92.6}Or_{0.1}$ | / | 4.9* | 13.8* | – | 5.24* | 32.30* |
| E | JaH 846[b] | OC3 | ≥3.2 | 13.5* | 622 | 77.2* | 560 ± 380 | | W2 | 0.6–38.1 | 9 | 1.9–35.2 | 3 | 0.5–4.4 | $An_{10.2}Or_{3.4}$* | 4* | 6.9* | 1.9* | – | 4.91* | 2.30* |
| F | NWA 12334[e] | LL6-an | 6 | 19.3* | 445 | 64.9* | – | – | / | 34.1 ± 0.3 | 8 | 27.5 | 1 | 2.4 | $An_{4.1}Or_{15.9}$ | 1 | 11.7* | – | 4.0* | 3.06* | 0.13* |
| F | NWA 12480[e] | Acapulcoite | – | – | – | – | – | – | minimal | – | – | 11.2 ± 0.2 | 4 | 2.1 ± 0.3 | $An_{16.6}Or_{4.4}$ | 2 | 6.0 | 12.0 | – | 5.08* | 36.6* |
| F | NWA 6592[n] | Lodranite | – | – | – | – | – | – | W1 | 16.0* | 2* | 6.9* | 1* | 46.0* | $An_{22.2}Or_{3.5}$* | 5* | 0.2 | 11.2 | – | 5.70* | / |

# Table 1

**Table 2**
Asteroids matching to meteorites.

| Petro/Spectral group | Meteorite | Classification | Qualitative Match (QM) | | | Highest Recurrent Match (HRM) | | | | | | Closest Matching asteroids (CMA) | | | Bus DeMeo taxonomy | | | Conclusion |
|---|---|---|---|---|---|---|---|---|---|---|---|---|---|---|---|---|---|---|
| | | | Powder | Raw | P.S. | Powder | *Score* | Raw | *Score* | P.S. | *Score* | Powder | Raw | P.S. | Powder | Raw | P.S. | |
| A/1 | Chwichiya 002 | C3.00-ung | C | B | B | Cg-Cgh-Ch-Xc-Xe | *1/1* | B | *3/6* | C-Cg-Cgh-Ch-B-Xc-Xe | *1/1* | Cg | B | B | Cgh-Cg-Xk-Xn | B | B | B-C-Cg |
| A/1 | NWA 11750 | C3-ung | C | ** | Xc | Cgh-Ch-B | *1/1* | B | *6/6* | ** | ** | Ch | B-C-Cgh | Ch | B | B | Cgh-Cg-Xk-Xn | B-Ch-Cgh-Cg |
| A/1 | NWA 12957 | C-ung | Cg | Cg | – | Cg-Cgh-Ch-B | *3/6* | Cg | *3/6* | – | – | B | Cg | – | Cgh-Cg-Xk-Xn | B | – | Cg |
| B/1 | EM 200 | C3 | C | C | C | Cgh-Ch-B-Xc-Xe | *1/1* | B-Ch | *3/6* | C-Cg-Cgh-Ch-Xk-B | *3/6* | Ch-Cg | Cg | Cg | C-B | B | Ch-Xk-Xn | C-Ch-Cg-B |
| B/2 | Aydar 003 | CM1/2 | Ch | – | Ch | Cg-Cgh-Ch | *3/3* | – | – | C-Cg-Cgh-Ch-Xk-B-Cb-Xc-Xe | *1/1* | B | – | Ch | C-Ch-Xk-Xc-Xe-Xn | – | B | Ch |
| B/2 | EM 100 | C2-ung | ** | ** | – | C | *1/6* | B | *1/6* | – | – | C | B | – | A-D | C-Ch-X-Xk-Xe-Xn | – | ? |
| B/2 | NWA 11086 | CM-an | ** | ** | ** | C-Cg-Xk | *3/6* | Xk | *2/6* | Cgh-Ch-Xk | *5/6* | C-Cg | Ch-Xk | Cgh | K-Xe | ** | ** | ? |
| B/2 | NWA 5958 | C2-ung | ** | B | – | Ch-Cgh-B | *4/6* | B | *1/6* | – | – | B-Ch | ** | – | K-Xe | K-Xe | – | Ch-B |
| A//23 | Acfer 094 | C2 | ** | ** | ** | Ch-Cgh-B | *3/6* | Ch-B | *3/6* | Xk-Cgh-Ch | *3/6* | B-Ch | B-Ch | B-Ch-Cgh | Xe-L | Xe-L | Xe-L | ? |
| C/3 | NWA 12474 | CR3.1 | ** | ** | – | B | *3/6* | B | *3/6* | – | – | B | B | – | B | B | – | ? |
| C/3 | Dho 2066 | C-ung | Xc | Xe-Xk | Xc | Ch-Cg-Cgh-Xk-Xe-Xc | *1/1* | Cgh-Ch-B-Xe-Xc | *1/1* | ** | ** | Xc | Ch | L | Xk-Xc-Xe-C-Ch-Xn | B | Cgh-Cg-Xk-Xn | Xc-Xe-Xk |
| C/3 | Kakangari | K | ** | ** | ** | Ch-Cgh-B | *3/6* | Xk-Cgh-Ch-B | *3/6* | C-Cgh-Xk-Cb-T-X-Xc-Xe | *1/1* | Cgh | B | T | Xe-L | B | Xe-L | Xe-Xk-Cgh-Ch |
| C/3 | LoV 200 | CH3 | Xe-Xk | Xe-Xk | Xc | Ch-Cg-Cgh-Xc-Xe | *1/1* | Xe-B-Ch-L | *1/1* | ** | ** | Xc | B | ** | Xk-Xc-Xe-C-Ch-Xn | B | A-D | Xe-Xk |
| B/3 | NWA 8781 | C-ung | ** | ** | – | Ch-Cgh-Cg | *3/6* | Ch | *6/6* | – | – | Xk | B-Ch | – | Xe-L | Xe-L | – | Xe-Xk |
| C/3 | SG 009 | Chond-ung | ** | ** | ** | Xk-Cgh-Ch | *6/6* | B-Cg | *2/6* | Xk-Cgh-Ch | *2/6* | Cg-Ch | B-Cg | Cg-Xk | Xe-L | Xe-L | Sv | Xk-Cg |
| D/4 | Awsserd | R4 | K | ** | K | K-S | *10/15* | Sr | *6/15* | K-S-Sq-Sr-Q | *10/15* | Sq-K | S-Sa-Sr | S | K-Xe | ** | K-Xe | K |
| D/4 | NWA 12472 | R3 | K | K | K | K-Sr-S | *10/15* | S | *4/15* | K-O-S-Sr-Sv | *6/15* | Sq-K | Q | Q | K-Xe | B | Xe-L | K |
| E/5 | LoV 051 | C3-ung | S-Sr-Sv | S-Sr-Sv | S-Sr-Sv | S-Sr | *6/15* | S-Sr-Sv | *6/15* | R-Sv | *6/15* | S-Sv | S | Sv | S | ** | S | S-Sr-Sv |
| E/5 | Coolidge | C4-ung | ** | – | S-Sr-Sq | S-Sr | *6/15-* | – | – | K-R-S-Sq | *3/15* | S | – | R/Sq | C | – | ** | S-Sr-Sq |
| E/5 | SaH 00177 | C3/4-ung | – | – | S-Sr-Sv | – | – | – | – | S | *9/15* | – | – | Sr | – | – | ** | S-Sr-Sv |
| F/5 | NWA 12334 | LL-an | S-Sr | ** | – | Q-K-Sq | *15/15* | S | *6/15* | – | – | Sq | S | – | K-Xe | B | – | S-Sr-Sq |
| E/6 | JaH 846 | OC3 | Sq-Sv | Sq-Sv | – | Sr | *15/15* | S-Sr | *10/15* | – | – | R | Sr | – | Sv | S | – | S-Sr-Sv |
| F/6 | Mulga (west) | C5/6-ung | – | – | Sa-Sv | – | – | – | – | K-Sq | *8/15* | – | – | Sq | – | – | | Sa-Sq |
| F/6 | NWA 12480 | Acapulcoite | Sv | Sv | Sv | V-R | *3/15* | V | *5/15* | R | *5/15* | R-V | V | Sv-V | V | V | Vw | V-Sv |
| F/6 | NWA 6592 | Lodranite | – | – | Sa | – | – | – | – | Sa-A-K | *1/15* | – | – | Sa | – | – | X-Xk-Xe-C-Xn | Sa |

**: no match; –-: no data; Chond-ung: chondrite ungrouped; P.S.: polished section; Petro group: petrographic group.

Table 2